\def\cm-2{$cm^{-2}$}

\def\ferg{ergs s$^{-1}$ cm$^{-2}$}

\def\kms{km s$^{-1}$}
\def\kmsMpc{km s$^{-1}$ Mpc$^{-1}$}
\def\femg{\ion{Fe}{2}/\ion{Mg}{2}}
\def\fe2{\ion{Fe}{2}}
\def\mg2{\ion{Mg}{2}}

\documentclass[12pt, preprint]{aastex}

\begin{document} 
 
\title{\ion{Fe}{2} EMISSION IN 14 LOW-REDSHIFT QUASARS: I - Observations} 

\author{Yumihiko Tsuzuki\altaffilmark{1}\altaffilmark{2}}
\author{Kimiaki Kawara\altaffilmark{2}}
\author{Yuzuru Yoshii\altaffilmark{2}} 
\author{Shinki Oyabu\altaffilmark{3}}
\author{Toshihiko Tanab\'{e}\altaffilmark{2}}
\author{Yoshiki Matsuoka\altaffilmark{2}}

\altaffiltext{1}{Institute for Cosmic Ray Research, University of Tokyo,
5-1-5, Kashiwanoha, Kashiwa, Chiba 277-8582, Japan; 
ytsuzuki@icrr.u-tokyo.ac.jp}
\altaffiltext{2}{Institute of Astronomy, University of Tokyo, 2-21-1, Osawa, Mitaka, Tokyo 181-0015, Japan}
\altaffiltext{3}{Institute of Space and Astronautical Science, Japan
Aerospace Exploration Agency, 3-1-1, Yoshinodai, Sagamihara, Kanagawa
229-8510, Japan}

\begin{abstract}

We present the spectra of 14 quasars with a wide coverage of rest wavelengths from 1000 to 7300 \AA. 
The redshift ranges from $z =$ 0.061 to 0.555 and the luminosity from $M_{B}$ = $-$22.69 to $-$26.32. 
These spectra of 
high quality result from combining {\it Hubble Space Telescope} spectra with those taken from ground-based telescopes.
We describe the procedure of generating the template spectrum of \ion{Fe}{2} line emission from the spectrum of a narrow-line Seyfert 1 galaxy I Zw 1 that covers two wavelength regions of 2200$-$3500 \AA\ and 4200$-$5600 \AA.  
Our template \fe2\ spectrum is semi-empirical in the sense that the synthetic spectrum calculated with the 
CLOUDY photoionization code is used to separate the \fe2\ emission from the \mg2 $\lambda$2798 line.
The procedure of measuring the strengths of \fe2\ emission lines is twofold; (1) subtracting the continuum components 
by fitting models of the power-law and Balmer continua in the continuum windows which are 
relatively free from line emissions, and (2) fitting models of the \fe2\ emission based on the \fe2\ template to the continuum-subtracted spectra. 
From 14 quasars including I Zw 1, we obtained the \fe2\ fluxes in five
 wavelength bands ($U1$ (2200$-$2660 \AA), $U2$ (2660$-$3000 \AA), $U3$
 (3000$-$3500 \AA), $O1$ (4400$-$4700 \AA), and $O2$ (5100$-$5600 \AA)),
 the total flux of Balmer continuum, and the fluxes of \mg2
 $\lambda$2798, H$\alpha$, and other emission lines, together with 
the full width at half maxima (FWHMs) of these lines. 
Regression analysis was performed by assuming a linear relation between
 any two of these quantities. Eight correlations were found with a
 confidence level higher than 99\%; (1) larger \ion{Mg}{2} FWHM for
 larger H$\alpha$ FWHM, (2) larger $\Gamma$ for fainter $M_B$, (3)
 smaller \ion{Mg}{2} FWHM for larger $\Gamma$, (4) larger \ion{Mg}{2}
 FWHM for smaller \fe2($O1$)/\ion{Mg}{2}, (5) larger $M_{BH}$ for
 smaller $\Gamma$, (6) larger $M_{BH}$ for smaller
 \fe2($O1$)/\ion{Mg}{2}, 
(7) larger [\ion{O}{3}]/H$\beta$ for larger \ion{Mg}{2} FWHM, and (8) larger \ion{Fe}{2}($O1$)/\ion{Mg}{2} for larger \ion{Fe}{2}($O1$)/\ion{Fe}{2}($U1$).
The fact that six of these eight are related to FWHM or $M_{BH}$ ($\propto$ FWHM$^2$) may imply that $M_{BH}$ is a fundamental quantity that controls $\Gamma$ or the spectral energy distribution (SED) of the incident continuum, which in turn controls the \fe2\ emission. 
Furthermore, it is worthy of noting that \ion{Fe}{2}($O1$)/\ion{Fe}{2}($U1$) is found to tightly correlate with \ion{Fe}{2}($O1$)/\ion{Mg}{2}, but not with \ion{Fe}{2}($U1$)/\ion{Mg}{2}. 

\end{abstract}

 
\section{INTRODUCTION} 

There is a growing interest in observing prominent \ion{Fe}{2} emission
lines in the spectra of active galactic nuclei (AGNs) to understand the
physics of clouds in the broad emission line regions (BELRs) and also to explore the \ion{Fe}{2} abundance as a function of cosmic time (e.g., Elston, Thompson, \& Hill 1994; Kawara et al. 1996; Dietrich et al. 2002, 2003; Iwamuro et al. 2002, 2004; Freudling, Corbin, \& Korista 2003; Maiolino et al. 2003). According to explosive nucleosynthesis, much of the iron is 
produced by Type Ia supernovae (SNe Ia), while $\alpha$ elements such as O and Mg come from Type II supernovae (SNe II). 
Because progenitors of SNe Ia are 
long-lived, accreting white dwarfs in binaries, while those of SNe II are short-lived massive stars, 
Fe enrichment delays relative to $\alpha$ elements. A lifetime of progenitors of SNe Ia is generally considered to be $\tau_{\rm Ia}\sim$ 1$-$2 Gyr \citep{hf, ytn, ytk}.
For a cosmology of $H_0 = 70$ \kmsMpc, $\Omega_M = 0.3$, and $\Omega_\lambda = 0.7$ \citep{ben}, such lifetime corresponds to the cosmic time at $z \sim$ 3.2$-$5.6. Therefore, hoping to detect a sudden break in the Fe/Mg abundance ratio at high redshift, various groups have measured the flux ratio of \ion{Fe}{2} emission relative to \ion{Mg}{2} $\lambda 2798$, \ion{Fe}{2}/\ion{Mg}{2}, in high-redshift quasars.

However, no clear break has been seen in a plot of \ion{Fe}{2}/\ion{Mg}{2} as a function of lookback time up to $z \sim 6$. If the Fe/Mg abundance ratio is reflected in the \ion{Fe}{2}/\ion{Mg}{2} flux ratio, no break in the plot implies that a lifetime of progenitors of SNe Ia should be much less than that generally considered. Accordingly, a significantly shorter lifetime of $\tau_{\rm Ia}\sim$ 0.2$-$0.6 Gyr is recently suggested \citep{ft, mr, gra04}, based on the binary prescription by \citet{mg} where $\tau_{\rm Ia}$ was said to be constrained by observations of binaries in the solar neighborhood. However, it is well known that the observed break of [$\alpha$/Fe] at [Fe/H] = $-$1 in the solar neighborhood is not explained by this prescription. 

The expected break of Fe/Mg at a certain high redshift would be obscured in 
\ion{Fe}{2}/\ion{Mg}{2} by other causes. 
First of all, the excitation mechanism 
of \ion{Fe}{2} emission lines has not been well understood. For example, 
although model simulations are often performed in the framework of 
photoionization \citep{sp, v03, b04}, it is sometimes pointed out that pure
photoionization models may not be sufficient to explain the spectrum of strong 
\ion{Fe}{2} emitters and a non radiative heating mechanism such as a shock 
has been suggested especially for the optical 
\ion{Fe}{2} emission (e.g. Collin \& Joly 2000). In addition, such model simulations imply that the 
relative iron abundance is only one parameter and furthermore not the most 
important one at all. Such effects of non-abundance factors include the 
spectral energy distribution (SED), strength of the radiation field, and the 
gas density of BELR clouds. Recently, \citet{v03} and \citet{b04} pointed out 
that a large microturbulence with $V_{turb} \geq$ 100 km s$^{-1}$ may be 
responsible for strong \ion{Fe}{2} emission. It is therefore important to 
explore the \femg\ flux ratio in the parameter space of non-abundance factors 
and to find the dependence of \femg\ on these factors. 

Another cause which might obscure the break of Fe/Mg would come from 
uncertainties in measuring the respective strengths of Fe II and Mg II. 
To accurately measure the \ion{Fe}{2} strength especially in the UV region, 
it is necessary to determine the contribution from the continuum emission 
which dominates the UV and optical spectra of quasars. This continuum 
subtraction becomes more accurate when a wider wavelength coverage of the 
spectrum is available. 

In this paper, we present the spectra of 14 quasars with a wide coverage of 
rest wavelengths from 1000 to 7300 \AA, and measure the relative line strength 
\femg\ accurately. A search for non-abundance effects in the \femg\ will be 
made at the end of this paper. A cosmology of $H_0$ = 70 \kmsMpc, 
$\Omega_M = 0.3$, and $\Omega_\lambda = 0.7$ will be used throughout.

\section{OBSERVATIONS} 

\subsection{Sample Selection}

To obtain the high quality spectra including the UV and optical \ion{Fe}{2} 
emission lines, we searched the {\it Hubble Space Telescope (HST)} 
archive for the spectra of low-redshift quasars listed in the \citet{vv98}
\footnote{According to the definition by V\'{e}ron-Cetty and V\'{e}ron, quasars are point sources with 
$M_B \leq -23$ for $H_0 = 50$ km s$^{-1}$ Mpc$^{-1}$ with $q_0=0$, approximately corresponding to those of $ z < 0.6$ that have $M_B \leq -22.35$ for the cosmology used in this paper.},
by using the following two criteria; (1) the spectrum should cover a wide range of rest wavelengths from 1200 to 6200 $\mbox{\AA}$, and (2) the quality should be good with a signal-to-noise ratio (SNR) of 20 or better. We found 14 quasars of $z \leq 0.6$ that pass the above criteria as of July 1999. 
For the optical and near-infrared parts of the spectra, we first looked at the {\it HST} archive and the literature. 
When no high-quality spectra are available, we have performed the optical and near-infrared spectroscopy. Table \ref{sample} gives a list of our quasar sample with $0.061 \le z \le 0.555$ and $-26.32 \le M_{B} \le -22.69$. 

It should be noted that I Zw 1, on the first row of the table, is the prototype narrow-line Seyfert 1 galaxy (NLSy1). The NLSy1 class of AGNs is characterized by the narrow-line profile having the full width at half maximum (FWHM) of H$\beta < 2000$ km s$^{-1}$ together with strong \ion{Fe}{2} emission lines, and furthermore tends to have steep soft X-ray spectra and strong IR emission (e.g., Sulentic, Marziani, \& Dultzin-Hacyan 2000; Laor et al. 1997a; Lipari 1994). 
I Zw 1 is a strong \ion{Fe}{2} emitter with average line width of
$\sim$ 900 km s$^{-1}$ \citep{vw} which is narrow enough to separate \ion{Fe}{2} emission from other lines such as \ion{C}{3}] and \ion{Mg}{2} in the UV region,and [\ion{O}{3}], H$\beta$, and H$\gamma$ in the optical region. In this paper, we obtain the UV/optical \ion{Fe}{2} template from the I Zw 1 spectrum, 
then fit it to the spectra of other quasars to measure their \ion{Fe}{2} emission fluxes.

\subsection{Spectra taken with {\it HST}}

As summarized in Table \ref{hst}, the UV spectra of all the quasars and optical/near-infrared spectra of four quasars have been taken with the Faint Object Spectrograph (FOS) and/or the Space Telescope Imaging Spectrograph (STIS) onboard the {\it HST}. The FOS has two Digicon detectors: `BLUE' and `AMBER (RED)'. `BLUE' was used for the shortest wavelength range of 1140$-$1606 \AA, and `AMBER (RED)' for the other longer wavelength range of 1590$-$8500 \AA. The resolution $\lambda/\Delta \lambda$ is about 1300 for both detectors. The STIS has two gratings, namely, G430L covering the UV/optical range of 2900$-$5700 \AA\ with a dispersion of $\Delta \lambda$ of 2.73 \AA/pixel, and G750L covering the optical/near-infrared range of 5240$-$10270 \AA\ with $\Delta \lambda$ of 4.92 \AA/pixel.    

For FOS observations, the raw spectra and the calibration files were retrieved 
from the archive. The Space Telescope Science Data Analysis System (STSDAS) on 
the IRAF\footnote{IRAF is distributed by the National Optical Astronomy
Observatories, which are operated by the association of Universities for
Research in Astronomy, Inc., under cooperative agreement with the
National Science Foundation.} was used for data 
reduction. The spectra were processed with the $calfos$ routines. To the 
spectra taken with the `BLUE' Digicon, the correction for the zero-point 
wavelength 
was applied by using the $foswcorr.cl$ script.  The resultant spectra for all 
the FOS wavelength ranges were combined into the single spectrum, and finally 
the $trebin$ script was applied for rebinning the original 0.2 \AA/pixel into 
2 \AA/pixel. For STIS spectra, the flux- and wavelength-calibrated spectra are 
already available in the achieve. We thus simply retrieved the spectra and 
combined them with FOS spectra without rebinning. In the overlapping region of {\it HST} 
spectra, no significant variations in continuum level were found 
except for I Zw 1. The {\it HST} spectra of I Zw 1 were taken six months apart, and 
the variation was recorded, which is consistent with the known variability 
of I Zw 1 \citep{vw}. The flux-scaling of {\it HST} spectra of this quasar will be 
discussed in section 2.5.

\subsection{Spectra Taken with Ground-based Facilities}

Table \ref{ground} lists 10 quasars whose optical spectra were not available 
in the {\it HST} archive.
Among these, the optical/near-infrared spectra of three quasars are in the 
literature and the Issac Newton Group (ING) archive, i.e., I Zw 1 in 
\citet{l97b}, QSO 0742+318 in \citet{wnw}, and B2 2201+31A in \citet{c97}. 
Optical/near-infrared spectra of the other seven quasars were newly observed 
using the Goldcam on the 2.1m telescope at the Kitt Peak National Observatory 
(KPNO) in 2001 May. These observations are labeled ``This work'' in Table 
\ref{ground}. Grating \#32 ($\Delta \lambda = $ 2.47 $\mbox{\AA}$/pixel) was used to cover the wavelength range of 4300$-$9200 \AA. The wavelength and flux calibrations were performed by using the HeNeAr lamp and 
observing two standard stars, namely, GD140 and HZ44 \citep{ma}. 

In addition, 
we made near-infrared spectroscopy of 3C 334.0 at $z = 0.56$, the highest 
redshift quasar in our sample, with the Cooled Infrared 
Spectrograph and Camera for OHS (CISCO) on the Subaru 8.2 m telescope on 2001 
May 9.  Total exposure time was 600 sec and the $zJ$ grism (0.87$-$1.39 $\mu$m 
with $\Delta \lambda$ of 5.83 $\mbox{\AA}$/pixel) was used.
OH lines and HD140385 (G2V) were used to calibrate the wavelength and
flux scales, respectively.

\subsection{Combining Spectra and De-reddening Galactic Extinction}

The wavelength- and flux-calibrated spectra in the UV, optical, and near-infrared regions were then combined into a single spectrum by equalizing the flux densities in the overlapping region of the spectra. We assumed that the UV flux density is the most accurate; the UV flux was fixed, while the optical and near-infrared fluxes were scaled. Before combining the spectra, the mean flux densities in the overlapping region of the ground-based spectra are 1.0$-$2.2 times smaller than those of the {\it HST} spectra, except for QSO 0742+318 whose mean flux density of the ground-based spectra is 1.3 times greater than those of the {\it HST} spectra. These differences in continuum level are expected from flux loss by the seeing effect and the intrinsic variability of AGNs. The combined spectra are plotted in the observer's frame without any reddening correction in Figure \ref{ObservedSpectra} \footnote{Throughout this paper, the spectra are plotted as wavelength $\lambda$ versus flux $\lambda$$F_{\lambda}$ aiming for the easy comparison between UV \ion{Fe}{2} and optical \ion{Fe}{2} in the energy unit.}.

The combined spectra were then de-reddened according to the extinction map of the Milky Way 
based on the far-infrared emission observed by {\it Infrared Astronomy
Satellite} ({\it IRAS}) and {\it Cosmic Background Explorer Satellite}
({\it COBE}) \citep{sfd}.
They tabulated the color excess $E_{B-V}$ for interstellar extinction with a 
resolution of 6.1$\arcmin$.
The accuracy of $E_{B-V}$ values is 16\%. As shown in Table 
\ref{sample}, our quasars have a range of $E_{B-V}$ = 0.01$-$0.06 
except for B2 2201+31A having $E_{B-V} = 0.12$. Because 
the extinction in magnitude at 2500 \AA\ around the strong \ion{Fe}{2} 
emission feature is given by $A_{2500} = 7.27E_{B-V}$ for the Milky Way 
dust with $A_V/E_{B-V} = 3.08$ \citep{p92}, the values of $E_{B-V} = 0.01, 
0.06,$ and 0.12 correspond to $A_{2500} = 0.07, 0.43,$ and 0.87, respectively, 
implying that the Galactic correction is significant for some quasars having 
large $E_{B-V}$.  
The combined spectra which have been corrected for the Galactic
extinction using the extinction curve of Milky Way by 
\citet{p92} are shown later in the left panels of Figure 
\ref{qso1}. In addition to the $B$ magnitude and $B-V $color derived from these
spectra, other wavelength properties are summarized in Table
\ref{multi}.

\subsection{Comments on the I Zw 1 Spectrum}

Because of its usefulness as a template for identifying the weak features in other AGNs and a benchmark for testing the models of the complex \ion{Fe}{2} emission, various groups have analyzed the UV/optical spectrum of I Zw 1. \citet{l97b} presented the spectrum of 1000$-$6000 \AA\ combining the {\it HST} FOS spectra with the optical spectrum taken from the KPNO 2.1m telescope. \citet{vw} analyzed the spectrum of 1000$-$3200 \AA\ made up from the {\it HST} FOS spectra only. \citet{vjv} constructed a synthetic \ion{Fe}{2} template in the wavelength range from 3500 to 7500 \AA\ by using the optical spectra observed at the William Herschel Telescope (WHT) and the Anglo-Australian Telescope (AAT). \citet{b04} used the spectrum of 1000$-$6000 \AA, which is similar to that by \citet{l97b}, to compare with the models of \ion{Fe}{2} emission. \citet{vw} derived an UV template for \ion{Fe}{2} and \ion{Fe}{3} emission, but their template may not be very accurate because their spectrum does not cover the Balmer edge at $\lambda = 3647$ \AA\ which is necessary to accurately estimate the contribution from the Balmer continuum to the baseline under \ion{Fe}{2} UV emission.

As shown in Table \ref{hst} and \ref{ground}, combining the {\it HST} spectra 
with two ground-based spectra covering 3183$-$4073 \AA\ and 4008$-$7172 \AA, 
we have produced an entire spectrum of 1140$-$7172 \AA. We used the same 
method of processing the {\it HST} spectra as done by \citet{vw}.
It should be noted that \citet{l97b} applied the old calibration
file to the G130H spectrum (1140$-$1606 \AA) that underestimates the flux 
by 9\% \citep{vw}. \citet{vw} scaled the G130H spectrum by a factor of 1.3 
to fit the G190H/G270H spectrum, otherwise the mean flux of the G130H 
spectrum is 1.3 times smaller than that of the G190H/G270H spectrum in the 
overlapping region of the spectra.
We applied the new calibration file to the G130H spectrum and scaled the 
flux by a factor of 1.3. 
Thus, our spectrum is identical to that by \citet{vw} shortward of 3200
\AA, but is extended to 7172 \AA.

\subsection{Intrinsic Extinction} 

The Galactic extinction in line of sight to I Zw 1 is estimated to be 
$E_{B-V} = 0.06$ from the {\it IRAS}/{\it COBE} far-infrared maps \citep{sfd}. 
The extinction can also be estimated from the strength of 
\ion{O}{1} $\lambda$8446 (from $3p^3P$ to $3s^3S^0$) and \ion{O}{1} $\lambda$1304 
($3s^3S^0$ to $2p^3P$). Because the lower level of the $\lambda$8446
transition is the upper level of $\lambda$1304, the photon 
flux ratio of $\lambda$8446/$\lambda$1304 is unity if no reddening is
present \citep{nd}. \citet{kk81} found that this ratio can increase 
from unity to 1.3 in their standard model because of the 
Balmer continuum absorption of $\lambda$1304 photons and production of 
$\lambda$8446 photons by collisional excitation ($3s^3S^0$ to $3p^3P$). 
Comparing the relative strengths in the three \ion{O}{1} lines of 
$\lambda$1304, $\lambda$8446, and $\lambda$11287, \citet{mot} concluded 
that the collisional processes are important to determine the relative 
strengths in the \ion{O}{1} lines. 

Carefully analyzing \ion{O}{1} $\lambda$1304 that is blended with the 
\ion{Si}{2} doublet lines of $\lambda$1304 and $\lambda$1309, 
\citet{l97b} deduced the \ion{O}{1} $\lambda1304$ flux to be $10.9
\pm 0.6 \times 10^{-14}$ ergs s$^{-1}$ cm$^{-2}$ in February 1994. 
\citet{pm} observed the \ion{O}{1} $\lambda$8446 flux as $8.5 \pm 0.85
\times 10^{-14}$ ergs s$^{-1}$ cm$^{-2}$ in September 1983, 
in good agreement with $8.83 \times 10^{-14}$ ergs s$^{-1}$ cm$^{-2}$ 
observed in October 1998 by \citet{ru}. 
These observations were made 11 years apart, so that 
the continuum and lines should have varied in brightness as indicated from 
the variation in continuum between the {\it HST} spectra. To estimate the line ratio, we assume two extreme cases; 
(1) the strength of the \ion{O}{1} lines is constant in time regardless of variations in continuum, 
and (2) the equivalent width of the \ion{O}{1} lines is constant.
In the case 1, by just taking a ratio of two measurements, we obtain the photon flux ratio $\lambda$8446/$\lambda$1304 = 5.05 $\pm$ 0.61.  
In the case 2, the line strength should be re-scaled according to the difference in continuum level in the overlapping region of the spectra. Because
\citet{l97b} applied the old calibration file
to the G130H spectrum which underestimates the flux by 9\% and did not 
multiply a factor of 1.3 to this spectrum, the $\lambda 1304$ line flux 
measured using the G130H spectrum should be multiplied by 1.4. 
The strength of the spectrum by \citet{pm} covering a range of 8000$-$10000 \AA\ is 1.2 times smaller than ours in the overlapping region, therefore their 
$\lambda 8446$ flux should be multiplied by a factor 1.2. 
The photon flux ratio is then $\lambda$8446/$\lambda$1304 = 4.33 $\pm$ 0.52. 
We thus obtain $\lambda$8446/$\lambda$1304 = 3.8$-$5.7. Assuming the intrinsic 
ratio of 1.3 implies $E_{1304-8446}$ = 1.45$-$1.89, where 
$E_{1304-8446}$ = $A_{1304} - A_{8446}$ and $A_{\lambda}$ is the extinction in 
magnitude at $\lambda$ (\AA). The Galactic extinction $E_{B-V}$ = 0.06 gives 
$E_{1304-8446}$ = 0.45. Thus, the rest of the color excess 
$E_{1304-8446}$ = 1.0$-$1.44 would be caused by the intrinsic extinction. Assuming the relation $E_{1304-8446} = 14.33 E_{B-V}$ for the Small Magellanic 
Could (SMC) extinction curve \citep{p92}, we obtain $E_{B-V}$ = 0.07$-$0.1 for 
the intrinsic extinction of I Zw 1. 

The intrinsic extinction may account for the red UV/optical spectrum of I Zw 1.  After de-reddening the Galactic extinction of $E_{B-V}$ = 0.06, the power-law index deduced from our spectrum of I Zw 1 is $\alpha=-1.19$ ($F_{\nu} \propto \nu^{\alpha}$), which is still much redder than the median value $\alpha = -0.32$ found in the Large Bright Quasar Survey (LBQS) sample by \citet{fra} and the mean value 
$\alpha = -0.44$ determined by \citet{vb} based on the Sloan Digital Sky 
Survey (SDSS) quasar sample. It is noted that 95\% of LBQS quasars have 
$\alpha \geq -1$.
Correcting the spectrum for the intrinsic SMC-like reddening of $ E_{B-V} = 0.09$, inferred from the \ion{O}{1} lines, 
gives $\alpha = -0.47$, approximately accounting for the difference in $\alpha$ between I Zw 1 and LBQS and SDSS quasars. An idea of the intrinsic extinction would be consistent with the large infrared luminosity in I Zw 1 relative to the the optical luminosity. As shown in Table \ref{multi}, I Zw 1 has a relatively large $L_{60\mu m}/L_V$ = 2.1, where $L_{60\mu m}$ and $L_V$ are the 60 $\mu$m and $V$-band luminosities as defined by $L_{60\mu m}/L_V$ = $\nu_{60\mu m}F_{60\mu m}/\nu_VF_{V}$. 

A quasar PG 1114+445 also has a red index of $\alpha = -1.04$, which is the second reddest spectrum next to I Zw 1.  However, there is no clear evidence that PG 1114+445 has a strong infrared excess; three quasars other than I Zw 1 have an infrared excess ($L_{60\mu m}/L_V > 1$) greater than PG 1114+445 
($L_{60\mu m}/L_V = 0.89$), while they are bluer ($\alpha < -1$) than PG 1114+445. Hence, it is 
not certain that the red spectrum of PG 1114+445 is attributed to the intrinsic extinction. To evaluate the effect of intrinsic extinction on the flux ratios, especially \ion{Fe}{2}/\ion{Mg}{2} flux ratio, we analyze the spectra of I Zw 1 and PG 1114+445, assuming two extreme cases of no intrinsic extinction and significant intrinsic extinction. 
In the case of significant intrinsic extinction, we assume the SMC-type
extinction of $E_{B-V}$ = 0.09 for both I Zw 1 and PG 1114+445 which
blues these quasars to $\alpha = -0.5$ and $\alpha = -0.3$, respectively. 
The effect of intrinsic extinction on various correlations concerning 
\ion{Fe}{2} emission will be discussed in section \ref{RA}. 

\section{I Zw 1 Templates for \ion{Fe}{2} Emission}

The procedure of deriving the \ion{Fe}{2} template spectrum is similar to those described in \citet{BG} for the optical template and in \citet{cb} and \citet{vw} for the UV template; (1) subtracting the power-law and Balmer continua simultaneously from the I Zw 1 spectrum; 
(2) removing the emission lines other than \ion{Fe}{2}; (3) generating two-parameter family of \ion{Fe}{2} spectra, defined by line width and line strength by fitting the Gaussian profiles to the residual spectrum.     

\subsection{Continuum Windows \label{Continuum Windows}}

The UV and optical spectra of quasars are dominated by the power-law and Balmer continua. Therefore, reliable measurements of strength of blended \ion{Fe}{2} emission lines crucially depend on accurate subtraction of these continua. The continuum flux levels within the individual continuum windows are generally determined by assuming zero contributions from emission lines there. 
However, according to our experience, this assumption sometimes results in poor fit of the continuum models between Ly$\alpha$ and \ion{C}{4} $\lambda 1549$ where the determined continuum level is significantly lower than the observed flux. To avoid such an unrealistic fit, it is necessary to evaluate the contributions from emission lines within the continuum windows.  We thus simulated BELR clouds in the framework of photoionization, by using the CLOUDY photoionization simulation code \citep{fer} combined with a 371-level Fe$^+$ model \citep{v99}. 
Our photoionization models have incident continuum shapes defined by
\begin{equation} 
f_{\nu} = 
\nu^{\alpha(UV)}\textrm{exp}(-h\nu/kT_{cut})\textrm{exp}(-kT_{IR}/h\nu) + a\nu^{\alpha(OX)}, \label{seq}
\end{equation}
where $\alpha(OX)=-\Gamma + 1$, $\alpha(UV)$ = $-$0.9$-$+0.3, $\Gamma$ = 2.0$-$3.5, $T_{cut}$ = 1.5 $\times 10^5$ K, and $kT_{IR}$ = 0.136 eV. The coefficient $a$ is adjusted to produce the optical to the X-ray spectral index $\alpha(OX) = -1.4$.\footnote{If the continuum could be of a single power-law, $\alpha(OX)$ would be described as $f_{\nu}$(2 keV)/$f_{\nu}$(2500 \AA) = 403.3$^{\alpha(OX)}$, where $\alpha(OX) = -1.4$ for typical AGNs.} The incident continuum illuminates a single BELR cloud having the gas density of $N_H = 10^{9}-10^{11}$ cm$^{-3}$, the ionizing parameter of $U = 10^{-3}-10^{-0.5}$, the microturbulence 
of $V_{turb}$ = 0$-$10 km s$^{-1}$, and the solar abundance. 
Our calculations were performed for 810 sets 
of parameters with a grid of points, 
($\alpha(UV)$, $\Gamma$,  $N_H$, $U$, $V_{turb}$) 
= (3, 6, 3, 5, 3), where $\alpha(UV)$ = 3 means that the calculations were 
made for three different values of $\alpha(UV)$. 
An example of our synthetic spectra is shown in the upper panel of Figure \ref{modelspec} (hereafter called the model-A spectrum). After applying the internal SMC-like extinction of $E_{B-V}$ = 0.13, this spectrum reasonably reproduces the observed spectrum of PG 1626+554 as shown in the lower panel of Figure \ref{modelspec} (hereafter called the model-B spectrum). In the following, we use these single-cloud models.\footnote{Any model consisting of a single BELR cloud oversimplifies the real BELR cloud system. In fact, our single-cloud models with some internal extinction (i.e., the model-B spectrum) poorly reproduce high-ionization lines, namely, \ion{He}{2} $\lambda$1640, \ion{C}{4} $\lambda$1549, \ion{C}{3}] $\lambda$1909, and \ion{He}{2} $\lambda$4686.  This suggests that two BELR clouds are at least needed; one for reproducing high-ionization lines, and the other for low-ionization lines. However, our single-cloud models are useful to estimate the continuum levels within the continuum windows and to carry out some preliminary comparison of the photoionization models with low-ionization lines such as \ion{Fe}{2} and \ion{Mg}{2}. This is justified by the fact that there are no quasar spectra whose observed fluxes are below the levels of the model continua, after the contributions from emission lines in Table \ref{cws} are taken into account.}
It should be noted that \citet{b04} used the CLOUDY combined with the same 
371-level Fe$^{+}$ model as used in this paper, and introduced 
a microturbulence $V_{turb} > 100$ km s$^{-1}$ to fit the observed UV \ion{Fe}{2} feature. On the other hand, our model can reproduce the UV \ion{Fe}{2} emission feature with a modest microturbulence of $V_{turb} = 0-10$ km s$^{-1}$. This difference comes from the choice of the UV bump cutoff temperature $T_{cut}$; we used $T_{cut} = 1.5 \times 10^{5}$ K, while \citet{b04} used a higher temperature 10$^{6}$ K, causing the excessive reduction of the \ion{Fe}{2} emissivity which then 
needs to be compensated by increasing microturbulence. Further discussion will 
exceed the scope of this paper, and will be given in a forthcoming paper.

The continuum windows are listed in Table \ref{cws} together with the range of the contribution from all emission lines to the total flux in 810 synthetic spectra. Note that we assumed the covering factor of 0.5. This factor was derived from the comparison between the observed spectrum of PG 1626$+$554 and the model-B spectrum presented in the lower panel in Figure \ref{modelspec}. As shown in Table \ref{cws}, there are no ideal continuum windows where the contribution of emission lines can be ignored. However, several good continuum windows exist in the wavelength region longward of 3000 \AA\ where the contribution from all emission lines are only 1 or 2\% of the total flux. In the wavelength region shortward of 1500 {\AA}, the contribution from emission lines are relatively large at a level of 4$-$9\%. Considering the importance of constraining the continuum levels in both sides of the \ion{Fe}{2} emission features, we decided to use the continuum window of CW1 (1320$-$1350 {\AA}) in addition to the other five windows of CW3, CW4, CW5, CW6, and CW7 in Table \ref{cws}.  Thus, a total of such six windows were used to fit the power-law and Balmer continua to the quasar spectrum. 

\subsection{Continuum Subtraction \label{Continuum Subtraction}}

The following procedure was used to subtract the continua from the quasar spectrum. First, the observed flux in the continuum window is decreased by taking into account the predicted contribution from non-continuum emission listed in column 4 of Table \ref{cws}; for example, the total flux in CW1 is decreased by 7\%. Then, the power-law and Balmer continua are simultaneously fitted to the fluxes thus decreased in the six continuum windows. Finally, the power-law and Balmer continua are subtracted from the original, observed spectrum, resulting in the continuum-subtracted spectrum consisting of line emission alone.

Our fitting models of the power-law and Balmer continua $F_{\nu}^{Cont}$ are in the form of
\begin{eqnarray} 
F_{\nu}^{Cont} & = & F_0[(\nu/\nu_0)^{\alpha(UV)} + aF_{\nu}^{Bac}], 
\label{cont} \end{eqnarray}
where $\nu_0$ is the frequency at 5700 \AA\ and $F_{\nu}^{Bac}$ is the 
empirical distribution of Balmer continuum by \citet{gr82}:  
\begin{equation}
F_{\nu}^{Bac} = B_{\nu}(T_{e})(1-e^{-\tau_{\nu}}) \quad  \textrm{for}
\quad \nu \ge \nu_{BE}, 
\end{equation} 
where $\tau_{\nu} = \tau_{BE}(\nu/\nu_{BE})^{-3}$, and $B_{\nu}(T_{e})$ is the Planck function at the electron temperature $T_{e}$. $\nu_{BE}$ and $\tau_{BE}$ are the frequency and optical depth at the Balmer edge at $\lambda$ = 3647 {\AA}, respectively. Equation (\ref{cont}) has five fitting parameters, $F_0$, $\alpha$(UV), $a$, $T_{e}$, and $\tau_{BE}$. 

The procedure of subtracting the continua is shown in Figure \ref{zw1spec}, plotting the spectra of 
I Zw 1 
for two cases without and with correction for possible intrinsic extinction, 
namely, zero intrinsic extinction ({\it left} panels) and SMC-like intrinsic 
extinction of $E_{B-V}$ = 0.09 inferred from the \ion{O}{1} line ratio
({\it right} panels). Note that the correction for Galactic extinction is applied to both cases. The top panels show the spectrum to which the model continua are fitted. The best-fit models are shown by the solid line for the power-law continuum and by the dashed line for the Balmer continuum. 
The middle and bottom panels show the power-law subtracted spectra in the UV and optical. The dashed line in the middle panels is the best-fit Balmer continuum. Note that the Balmer continuum is zero 
longward of the Balmer edge at 3647 {\AA}. 
The spectra plotted in the middle and bottom panels are dominated by \ion{Fe}{2} emission in the UV and optical except for the Balmer continuum,
\ion{Mg}{2} $\lambda2798$, H$\gamma$, H$\beta$, 
[\ion{O}{3}] $\lambda$$\lambda$4959,5007, and [\ion{Fe}{2}] $\lambda$$\lambda$5158,5269 (see Figure \ref{zw1fe2}). 

The spectrum corrected for the intrinsic extinction ({\it right} panels) has the UV and optical \ion{Fe}{2} fluxes 
and the \ion{Mg}{2} line flux which are 1.59$-$2.09, 1.08$-$1.13, and 1.62 times greater than those deduced from 
the spectrum without intrinsic extinction ({\it left} panels), respectively. Thus, the line flux ratios, 
\ion{Fe}{2}(UV)/\ion{Mg}{2} and \ion{Fe}{2}(opt)/\ion{Mg}{2}, after corrected for the intrinsic extinction, are 
0.98$-$1.29 and 0.67$-$0.70 times those without intrinsic extinction, respectively.  It is therefore concluded that 
the error associated with the cases with and without intrinsic extinction is 30\% at most in \ion{Fe}{2}(UV, opt)/\ion{Mg}{2}. 
This error is much smaller than that in the Balmer continuum flux; the flux corrected for the intrinsic extinction 
is almost eight times smaller than that without intrinsic extinction (see Table \ref{lines1}).

\subsection{Template \ion{Fe}{2} Spectrum and Removal of Non-\ion{Fe}{2} Emission Lines}

The UV and optical spectra of I Zw 1, after subtracting the power-law and Balmer continua and correcting for the Galactic extinction only, are plotted in Figure \ref{zw1fe2}. The solid lines show the \ion{Fe}{2} spectra, while the dotted lines show the contributions of \ion{Mg}{2} $\lambda2798$ and other optical emission lines such as H$\gamma$, H$\beta$, [\ion{O}{3}] $\lambda$$\lambda$4959,5007, two [\ion{Fe}{2}] lines and one \ion{Fe}{2}] line which are blended near 5158 {\AA} (Moore's (1972) multiplet numbers of 18F $\lambda$5158, 19F $\lambda$5158, and 35 $\lambda$5161), and three [\ion{Fe}{2}] lines blended near 5269 \AA\ (19F $\lambda$5262 and 18F $\lambda$$\lambda$5269,5273). As shown in the left panel of Figure \ref{zw1ha}, the observed H$\alpha$ line profile is well fitted by two Gaussian components with different FWHMs of 690 and 2700 km s$^{-1}$ and the peak height ratio of 1:0.4. The 690 km s$^{-1}$ component is blueshifted relative to the 2700 km s$^{-1}$ component by 150 \kms. The broad component of \ion{Mg}{2} $\lambda2798$ with a FWHM of 5720 km s$^{-1}$ reported by \citet{vw} has not been confirmed in our analysis of the H$\alpha$ profile. In the right panel of Figure \ref{zw1ha}, the H$\alpha$ template profile composed of these two components is compared with the spectrum centered at H$\beta$. Differences between the observed spectrum and the H$\alpha$ template could be mostly attributed to broad features of \ion{Fe}{2} emission. 
The H$\alpha$ template is scaled in flux and shifted in wavelength to measure 
the contributions of the other broad emission lines, namely, 
H$\gamma$, H$\beta$, and \ion{Mg}{2} $\lambda$2798. For the Balmer lines, 
H$\gamma$, and H$\beta$, the relative strength of the two Gaussian 
components was fixed to that (1:0.4 for the peak height ratio) of 
the H$\alpha$ profile, while 
for the \ion{Mg}{2} $\lambda$2798 line, the relative strength is varied and 
determined to be 1:0.23 by fitting as described in the later paragraph in 
this sub-section.
The H$\alpha$ template cannot be
applied to the forbidden and semi-forbidden lines, namely, [\ion{O}{3}] 
$\lambda$$\lambda$4959,5007, [\ion{Fe}{2}], and \ion{Fe}{2}]. The FWHM 
measurements of [\ion{O}{3}] $\lambda$$\lambda$4959,5007 before subtracting 
the contribution of the \ion{Fe}{2} emission, are 910 and 1360 km s$^{-1}$, 
respectively, and are broader than the real FWHM values due 
to broadening by the broad \ion{Fe}{2} emission features.\footnote{In fact, 
the real FWHM value of [\ion{O}{3}] $\lambda$$\lambda$4959,5007, 
after subtracting the contribution of the \ion{Fe}{2} emission, is 640 $\pm$ 60
km s$^{-1}$ in I Zw 1 as given in Table \ref{tabFWHM}. 
For the other quasars, the FWHMs of [\ion{O}{3}] and 
the fluxes were also measured 
by using the same method, i.e., fitting a single-Gaussian component to 
the \ion{Fe}{2}-subtracted spectrum.
As given in Table \ref{tabFWHM}, the line widths of [\ion{O}{3}] are significantly 
narrower than those of \ion{Mg}{2} and H$\alpha$, confirming the 
narrow emission line region (NELR) clouds as the major source for [\ion{O}{3}] emission in 
quasars.} 
We therefore assumed the FWHM of the narrow 
component, 690 km s$^{-1}$, to be the FWHMs of these lines.
The relative intensities of [\ion{O}{3}] $\lambda$$\lambda$4959,5007 were 
assumed to be 1:3. For measuring the contributions of the [\ion{Fe}{2}] and 
\ion{Fe}{2}] lines, the line strengths relative to nearby \ion{Fe}{2} 
emission were fixed to those measured by \citet{vjv}.  

The subtractions of the forbidden and semi-forbidden lines may not be accurate
due to the large uncertainties of their FWHMs. 
Nonetheless, the errors associated 
with the subtraction of these lines is estimated to be very small and can be 
ignored for the following reasons; (1) two \ion{Fe}{2} templates covering 
two wavelength regions of 4400$-$4700 \AA\ and 5100$-$5600 \AA\ are used to 
measure the strengths of optical \ion{Fe}{2} emission, and none of them cover 
the wavelengths of the [\ion{O}{3}] lines, and (2) the contributions of 
the [\ion{Fe}{2}] and \ion{Fe}{2}] lines are very small when compared with 
the total strengths of the \ion{Fe}{2} emission in 4400$-$4700 \AA\ and 
5100$-$5600 \AA, i.e., 2$-$3\% and 3$-$4\%, respectively. It is noted that 
the wavelengths of H$\gamma$ and H$\beta$ are out of the range of optical 
templates.

\ion{Mg}{2} $\lambda$2798 is the $3s^2S-3p^2P^0_{1/2,3/2}$ resonance doublet 
at 2795.5 and 2802.7 {\AA}. This 2795.5/2802.7 doublet ratio changes from 
2/1 to 1/1 for an entirely thermalized gas \citep{l97b}. 
Because \ion{Mg}{2} $\lambda$2798 heavily blends with \ion{Fe}{2} emission 
lines, it is difficult to deduce the spectral feature of \ion{Fe}{2}
emission around \ion{Mg}{2} $\lambda$2798 from 
the observed spectrum. The choice of the \ion{Fe}{2} feature under 
the \ion{Mg}{2} line profile
would little alter the integrated flux of the broad \ion{Fe}{2} feature. 
However, it can significantly affect the \ion{Mg}{2} line flux. 
To determine the \ion{Fe}{2} emission level obscured by the \ion{Mg}{2} line 
profile, we take a 
semi-empirical approach with two assumptions: (1) \ion{Mg}{2} $\lambda$2798 
is purely made up with 
the $3s^2S-3p^2P^0_{1/2,3/2}$ resonance doublet, 
each of which has a line profile of the H$\alpha$ 
template, and (2) the spectral shape of \ion{Fe}{2} is the same as 
that in the model spectrum shown in Figure \ref{modelspec}. 

The model-A spectrum of \ion{Fe}{2} emission 
around \ion{Mg}{2} $\lambda$2798 is overplotted on
the I Zw 1 spectrum in the top panel of Figure \ref{zw1mg}, 
after convolved with a 690 \kms\ FWHM profile.  
The shaded area is a range allowed by 
synthetic \ion{Fe}{2} spectra 
in the parameter space used to study the continuum windows 
in section \ref{Continuum Windows}, 
where the synthetic spectra are normalized at 2775 \AA. 
To analyze the contribution of 
\ion{Mg}{2} $\lambda$2798, the \ion{Fe}{2} model-A spectrum was subtracted 
from the I Zw 1 spectrum and the resultant spectrum was fitted by 
the H$\alpha$ template with small modifications. 
Our \ion{Mg}{2} $\lambda$2798 models purely consists of the resonance 
doublet, and each line of 
the doublet has a line profile of the H$\alpha$ template. The relative 
strength of the doublet, the peak height ratio of the Gaussian 
component, and the relative velocity of the 2700 km s$^{-1}$ component to the 690 km s$^{-1}$ component are set to be free parameters. The best fit is shown in the middle panel of Figure \ref{zw1mg}, for the case that the relative strength of the doublet is 1.2:1, the peak height ratio of the Gaussian component is 1:0.23, and the relative velocity between the 2700 and 690 km s$^{-1}$ components is zero. 
Our best fit is in good agreements with \citet{l97b} who obtained 1.2:1 for relative doublet strengths by fitting the H$\alpha$ profile. When compared to \citet{l97b}, 
our fit is significantly improved especially over the both wings of the line profile with no need of a very broad 5720 km s$^{-1}$ component. 
This good fit would support our approach using model spectra to define the baseline of \ion{Mg}{2}$\lambda$2798.  The resultant semi-empirical \ion{Fe}{2} template spectrum is overplotted on the I Zw 1 spectrum in the bottom panel of 
Figure \ref{zw1mg}. This template spectrum is the difference of I Zw 1 spectrum minus the fit of \ion{Mg}{2}$\lambda$2798. 

In our approach, the flux of \ion{Mg}{2} $\lambda$2798 is the sum of the 
residual in the region of 2770$-$2830 \AA\ after subtracting 
the \ion{Fe}{2} template from the I Zw 1 spectrum.\footnote{
Once the semi-empirical \ion{Fe}{2} template spectrum is obtained from
the multiple Gaussian component analysis, we measured the line widths and 
strengths by performing the single Gaussian fitting to the \ion{Fe}{2} 
template-subtracted spectra. For example, as shown in 
Table \ref{tabFWHM}, the FWHM in the single Gaussian 
approximation is 1660 $\pm$ 10 km s$^{-1}$ in I Zw 1 for \ion{Mg}{2} 
$\lambda$2798 
and 1490 $\pm$ 20 km s$^{-1}$ for the H$\alpha$. It is noted that the 
single Gaussian component fitting was used because the multiple component 
fitting is only practical to exceptional high-quality spectrum like I Zw 1.}
The flux of \ion{Mg}{2} $\lambda$2798 thus measured is 
61.1 $\times 10^{-14}$ ergs s$^{-1}$ cm$^{-2}$. If the baseline of \ion{Mg}{2}
$\lambda$2798 is a straight line between the minimum points 
at 2771 and 2818 {\AA}, the \ion{Mg}{2} 
flux becomes 10\% smaller, while the baseline is defined by 
the minimum points allowed for our models, 
the flux becomes 8\% greater. Thus, the choice of the baseline is 
associated with a 10\% error in the \ion{Mg}{2} flux, 
which should be regarded as a typical size of the systematic error. 
Our \ion{Mg}{2} $\lambda$2798 flux is in between the other two values reported
by \citet{l97b} and \citet{vw}. 
The former is 52.1 $\times 10^{-14}$ ergs s$^{-1}$ cm$^{-2}$ (85 \% of ours), 
and the latter is 94.3 $\times 10^{-14}$ ergs s$^{-1}$ cm$^{-2}$ (154 \% 
of ours).  Difference in the \ion{Mg}{2} $\lambda$2798 flux between these 
authors is not very small, therefore a care is needed when comparing the 
fluxes taken from different sources.

Our \ion{Fe}{2} template spectrum is compared with that by \citet{vw} in the 
UV in the upper panel in Figure \ref{zw1comp} and that by \citet{vjv} 
in the optical in the lower panel, along with the model-A spectrum and the 
other synthetic spectrum. The template by \citet{vw} and 
the two synthetic spectra are scaled in flux, 
in such a way that the integrated fluxes match with our template spectrum 
between 2200 and 3100 \AA. Because the optical template by \citet{vjv} is 
given in an arbitrary units, it is scaled to match with our template spectrum 
in 4400$-$4700 \AA\ and 5100$-$5600 \AA. As a result of flux-scaling,
the original UV template spectrum by \citet{vw} was multiplied by a factor 
of 1.4.
The scaling of 1.4 in UV flux can mostly be attributed to the 
choice of the continuum level. They used a pure power-law continuum model to 
fit the I Zw 1 spectrum, and ignored the contribution from the Balmer 
continuum because their I Zw 1 spectrum does not cover wavelengths around 
the edge of the Balmer continuum which is used to measure the strength of 
the Balmer continuum. The power-law continuum fitting was made using the 
two continuum windows at 1675$-$1690 \AA\ and 3007$-$3027 \AA. As a result, 
their continuum level is significantly higher than ours as can be seen in 
the left panels of Figure \ref{zw1spec}, which in turn decreases their 
\ion{Fe}{2} flux. The magnification by a factor of 1.4 makes their template 
spectrum more spiky than ours. Nonetheless, there is a good agreement 
between the two template spectra except in the region around 2800 \AA. 
In the optical, there is a good agreement between the two template spectra 
except in the region around 4900 \AA. The disagreement around 4900 \AA\ may 
be caused by their H$\beta$ $\lambda$4861 profile. 
\citet{vjv} fitted H$\beta$ using the line profile comprised of a Lorentzian
component with a FWHM of 1100 km s$^{-1}$ and Gaussian component with a FWHM 
of 5600 km s$^{-1}$. As already discussed and shown in Figure \ref{zw1ha}, 
such a very broad component with a FWHM of 5600 km s$^{-1}$ is not confirmed 
in our analysis of the H$\alpha$ profile. 

Two synthetic spectra in the framework of photoionization are compared with 
observations; one is based on model-A and the other is a spectrum in 
a low-density cloud. The synthetic spectra were convolved with a line profile 
of a FWHM of 1660 km s$^{-1}$ that is the FWHM of 
\ion{Mg}{2} $\lambda$2798 in the single Gaussian form as discussed 
in footnote 7. 
As discussed in section 3, model-A is used to specify the 
contributions of emission lines to the spectra of normal quasars and 
has $N_H = 10^{10} $ cm$^{-3}$ with $U = 10^{-1}$ and 
$V_{turb}$ = 5 km s$^{-1}$. The low-density cloud model has
$N_H = 10^{7} $ cm$^{-3}$ with $U = 10^{-2}$ and $V_{turb}$ = 0 km s$^{-1}$ 
with other parameters identical to model-A.
It is generally considered that \ion{Fe}{2} emission arises in BELR clouds 
with a high gas density of $N_H \ge 10^{9.5} $ cm$^{-3}$ 
(e.g., Verner et al. 1999; Sigut \& Pradhan 2003). In this sense, model-A is 
more conventional than the low-density cloud (LDC) model.
Compared to the template spectrum, the model-A spectrum has an excess in 
flux around 2400 {\AA} while a deficiency longward of 2900 {\AA}. 
In the optical, the model-A spectrum fails to produce the \ion{Fe}{2} emission
by a factor of 10. This is a major reason that a non radiative heating 
mechanism, such as a shock model (e.g. Collin \& Joly 2000), is considered 
to be necessary to account for strong optical \ion{Fe}{2} emitters like I Zw 1.
However, it is interesting that the LDC model reproduces the strong optical 
\ion{Fe}{2} emission, although there is a severe deficiency around 4600 {\AA}.
Further investigation is required if photoionization can reproduce the 
strong optical \ion{Fe}{2} emission.  

\section{Application of the \ion{Fe}{2} Template Spectrum}

\subsection{Fitting to Quasar Spectra}

To apply the \ion{Fe}{2} template spectrum obtained from the NLSy1 I Zw 1 to 
other quasars, broadening of the template spectrum is needed to match with 
the line width of the quasar spectrum in consideration. We use the similar 
broadening method used by \citet{BG} and \citet{vw}; 
broadening the template using a relation, 
\begin{eqnarray} 
\textrm{FWHM}(\textrm{QSO})^{2} & = & \textrm{FWHM}(\textrm{convolution})^{2} + \textrm{FWHM}(\textrm{I Zw 1})^{2},  
\end{eqnarray}
where FWHM(QSO), FWHM(I Zw 1),\footnote{We used FWHM(I Zw 1) = 690 km s$^{-1}$, because 
it yields slightly better fit than that with FWHM(I Zw 1) = 1660 km s$^{-1}$ measured  
by fitting a single Gaussian to \ion{Mg}{2}. 690 km s$^{-1}$ is comparable to that of 
[\ion{O}{3}] FWHM (640 km s$^{-1}$) in I Zw 1, and [\ion{O}{3}] is generally considered 
to originate in NELR clouds. However, it is unlikely that a significant fraction of 
\ion{Fe}{2} emission comes from NELR clouds. As shown in panel $v$ in Figure 
\ref{correlations}, the \ion{Mg}{2} line profiles in other quasars are significantly 
broader than [\ion{O}{3}] profiles. Thus, we think that \ion{Fe}{2} emission is dominated 
by emission in BELR clouds.
}
and FWHM(convolution) are used to represent the line 
width of the target quasar, that of I Zw 1, and the width of additional convolution
applied to the I Zw 1 spectrum (i.e., broadening). 
The actual sequence of broadening would perform as follows: (a) 
the FWHM of \ion{Mg}{2} is measured in the original spectrum containing 
the \ion{Fe}{2} emission features; (b) this initial FWHM value is 
used to broaden the I Zw 1 template spectrum; 
(c) the broadened template is used to subtract the \ion{Fe}{2} emission 
features from the original spectrum; (d) the final FWHM value is determined 
by fitting a single Gaussian component to the \ion{Fe}{2}-subtracted spectrum;
(e) the final FWHM value is used to broaden the I Zw 1 template spectrum, and 
the \ion{Fe}{2} emission strengths is re-measured for the final results.
According to our experience, using the initial FWHM value of \ion{Mg}{2} is 
good enough to have accurate strengths of the \ion{Mg}{2} and \ion{Fe}{2} 
emission lines; the \ion{Mg}{2} and \ion{Fe}{2} fluxes measured using the 
initial FWHM value differ only $<$ 1\% typically or 3\% at most from those
measured using the final FWHM value. 

If the I Zw 1 spectrum is similar to those of other quasars, the \ion{Fe}{2} 
template spectrum could be fitted well simultaneously from the UV to the optical, 
using a single scaling in flux. However, I Zw 1 is a peculiar 
quasar as classified NLSy1 with extremely strong \ion{Fe}{2} emission, and the shape of the \ion{Fe}{2} 
spectrum 
would differ from other quasars. To flexibly cope with the possible variety of \fe2 spectrum shapes from 
quasar 
to quasar, we subdivide the template spectrum into five segments, and allow five separate flux-scalings for template fitting, i.e., one independent flux-scaling for each template segment. As shown by thick lines 
in Figure \ref{zw1fe2}, these segments correspond to the wavelength bands called $U1$ (2200$-$2660 \AA), $U2$ (2660$-$3000 \AA), $U3$ (3000$-$3500 \AA), $O1$ (4400$-$4700 \AA), and $O2$ (5100$-$5600 \AA). 
Fitting of the \ion{Fe}{2} emission is only performed within these bands. The fitting function of our \ion{Fe}{2} emission models is of the form:
\begin{eqnarray} 
F_{\nu}^{Fe} & = & \ \ \ F_{\nu}^{Fe}(U1)    + b_1F_{\nu}^{Fe}(U2) + b_2F_{\nu}^{Fe}(U3) \nonumber \\
             & + & b_3F_{\nu}^{Fe}(O1) +  b_4F_{\nu}^{Fe}(O2),
\label{fe2model} \end{eqnarray}
where $F_{\nu}^{Fe}(X)$ is the \ion{Fe}{2} template spectrum segment for 
the wavelength band of $X=U1, U2, .., O2$.  The coefficients $b_i$ are the flux-scaling parameters used to measure the relative strength of the \ion{Fe}{2} emission in the corresponding wavelength band. 
 
The model spectrum consisting of the power-law and Balmer continua and the \ion{Fe}{2} emission is therefore given by
\begin{eqnarray} 
F_{\nu}      & = & F_0[(\nu/\nu_0)^{\alpha(UV)} + aF_{\nu}^{Bac} + bF_{\nu}^{Fe}].
\label{allmodel} \end{eqnarray}
The best-fit models of this form are compared with the original spectra of 14 quasars in Figure \ref{qso1}. All the original spectra have been corrected for the Galactic extinction. The panels are sorted by $M_B$ with the lowest luminosity at the top. For two quasars with red spectral index, I Zw 1 and PG 1114+445, an alternative case is also shown such that the intrinsic SMC-like extinction of $E_{B-V}$ = 0.09 has been applied together with the Galactic extinction. The fits are summarized in Table \ref{fitparameter}. 

The lines of H$\beta$, H$\gamma$, and [\ion{O}{3}] as well as \mg2 were measured by fitting a Gaussian
\footnote{To measure the fluxes of the H$\beta$, H$\gamma$, and H$\delta$, a Gaussian profile derived 
from the H$\alpha$ spectrum by fitting a single Gaussian was used. For quasars without H$\alpha$ spectra, 
a Gaussian profile which reproduces both H$\beta$ and H$\gamma$ well was defined.}
 after subtracting the best-fit \ion{Fe}{2} emission model. It is noted that fitting was not performed 
in 4200$-$4400 \AA\ and 4700$-$5100 \AA\ where H$\beta$, H$\gamma$, and [\ion{O}{3}] are located. Instead, 
it is assumed that the spectra of 4200$-$4400 \AA\ and 4700$-$5100 \AA\ have the same strength as 
$b_3F_{\nu}^{Fe}(O1)$ and $b_4F_{\nu}^{Fe}(O2)$, respectively, at their boundaries. Fluxes of lines 
shortward of 2000 \AA\ were measured by fitting a Gaussian to the continuum-subtracted spectrum. Table 
\ref{tabFWHM} lists the FWHMs of \mg2, H$\alpha$, and [\ion{O}{3}], and the ratios between the flux of 
\ion{Mg}{2} and H$\alpha$ measured by fitting the two Gaussian components and that by fitting a single 
Gaussian component. Tables \ref{lines1} lists the fluxes 
of emission lines and the Balmer continuum as well as the flux at 1450 \AA. It should be kept in mind that 
the cases with and without intrinsic extinction can alter the line ratios significantly; for example, the 
correction for the intrinsic SMC-like extinction of $E_{B-V}$ = 0.09 increases the 
\ion{Fe}{2}($U1$)/\ion{Mg}{2} ratio by a factor of 1.3 and decreases \ion{Fe}{2}($O1$)/\ion{Fe}{2}($U1$) by 
a factor of two.  Furthermore, the Balmer continuum flux is decreased by a factor of eight. 

\subsection{Comparison with Previous Work}

We start with relating our \ion{Fe}{2} bands to those previously used in the literature. 
Optical \ion{Fe}{2} line emission, especially \ion{Fe}{2} $\lambda$4570 and $\lambda$5250 (referred to by 
\citet{phil} as the $\lambda$5190,5320 blend) has been observed in many 
quasars (e.g., Boroson \& Green 1992; Sulentic, Marziani, \&
Dultzin-Hacyan 2000). \ion{Fe}{2} $\lambda$4570 is a blend of
multiplets 37, 38, and 43 in the wavelength range between H$\gamma$ 
and H$\beta$, and \ion{Fe}{2} $\lambda$5250 is a blend of multiplets 42, 48, 49, and 55 in the wavelength range of 5100$-$5600 \AA\ \citep{phil}. In fact, the 
integrated \ion{Fe}{2} flux $F(O1)$ in the $O1$ band and similarly $F(O2)$ in  the $O2$ band are equivalent to the integrated fluxes usually used such as
\begin{eqnarray} 
F(\textrm{Fe II } \lambda 4570) & = & F(O1) \nonumber \\
F(\textrm{Fe II } \lambda 5250) & = & F(O2) .
\label{opt_trans} \end{eqnarray}
 
Prior to this work, the largest sample of quasar spectra with \ion{Fe}{2} emission observed both in the UV and optical is given by \citet{wnw}. They analyzed seven low-redshift quasars with a wide wavelength coverage from 1800 to 5500 \AA, and measured the \ion{Mg}{2} flux and the integrated \ion{Fe}{2} fluxes in 2000$-$3000 \AA, 3000$-$3500 \AA, and 3500$-$6000 \AA. Two quasars in their sample, QSO 0742+318 and 3C 273, are commonly included in our sample. 
Because our \ion{Fe}{2} bands are different from theirs, transformation equations between the two different systems are needed.  Figure \ref{cumulative_fe2} shows the cumulative \ion{Fe}{2} flux as a function of wavelength, derived from the best-fit \ion{Fe}{2} templates for I Zw 1 and PG 1626+554. We then find the following transformation equations:
\begin{eqnarray} 
F(2000-3000 \hspace{1mm} \mbox{\AA}) & = & 1.064 (\pm 0.005) \times [F(U1) + F(U2)]  \nonumber \\
F(3000-3500 \hspace{1mm} \mbox{\AA}) & = & F(U3) \nonumber \\
F(3500-6000 \hspace{1mm} \mbox{\AA}) & = & 1.997 (\pm 0.001) \times [F(O1) + F(O2)], \nonumber \\
                                     &   &  
\label{fe2transfer} \end{eqnarray}
where $F(X)$ are the integrated \ion{Fe}{2} fluxes in the wavelength range or 
band $X$.

The measurements of various emission fluxes in QSO 0742+318 and 3C273 by \citet{wnw} are compared with our measurements in Table \ref{compWNW}. No correction for the intrinsic extinction was made in both measurements. 
While their optical line strengths agree well with ours within an
accuracy of 20\%, those of UV lines shortward of 3000 \AA\ and the
Balmer continuum largely differ from each other and the difference is as
large as a factor of 2 to 3.  The reason for such a large difference
is not clear, but may possibly be because (1) the {\it International
Ultraviolet Explorer} UV spectra used by \citet{wnw} are not so good as expected, and/or because (2) the UV continuum levels estimated by \citet{wnw} are poorly determined due to their limited wavelength coverage in the UV. In fact, 
their UV continuum levels are 10\% smaller than ours, which increases their \ion{Fe}{2}(2000$-$3000 \AA) strengths by this factor when compared to ours.

\section{Correlations of \ion{Fe}{2} emission with other spectral
 properties \label{RA}} 

Large observational efforts have been devoted to searching for correlations between \fe2\ emission and 
other spectral properties. Such correlation studies were mostly made against
\fe2 $\lambda$4570, which is practically identical to our \fe2 ($O1$). 
This work should therefore be regarded as extension of previous studies by including UV \fe2\ emission. In the following, we simply show various correlations, although their statistical significance is limited because of our sample consisting of only 14 quasars. The non-abundance effects in the \ion{Fe}{2}/\ion{Mg}{2} flux ratio and related physical processes in BELR clouds will be discussed in a forthcoming paper. 
 
Various $X$ versus $Y$ diagrams are shown in Figure \ref{correlations}, which are useful to study the non-abundance effects in \fe2\ emission. Regression analysis was performed to derive a linear relation ($Y = A + BX$), weighted by individual standard deviations of two variables $X$ and $Y$. The results are summarized in Table \ref{rank}, showing the panel name identifying the diagram in Figure \ref{correlations}, the variables $X$ and $Y$, the sample number $N$, the intercept $A$, and the slope $B$, the linear-correlation coefficient $r$, and the confidence level. 

There are well-known correlations, namely, larger \fe2 ($O1$)/H$\beta$ for 
larger soft X-ray photon index $\Gamma$ \citep{wan, law, l97a}, for smaller 
FWHM of permitted lines \citep{ZO, ZK, law}, and for weaker 
[\ion{O}{3}]/H$\beta$ \citep{BG}. It is expected that these correlations are 
also seen for \fe2 ($O1$)/\ion{Mg}{2}, because both H$\beta$ and \ion{Mg}{2} are 
collisionally excited in the partially ionized region in the BELR clouds \citep{kk81}. In fact, 
these correlations in our diagrams give a confidence level of (panel $g$) 
97.9\%, (panel $k$) 99.3\%, and (panel $s$) 94.2\%, respectively. 

As shown in Table \ref{rank}, there are eight correlations with a confidence 
level of 99\% or higher, i.e., 
(1) (panel $a$) larger \ion{Mg}{2} FWHM for larger H$\alpha$ FWHM, 
(2) (panel $e$) larger $\Gamma$ for fainter $M_B$, 
(3) (panel $i$) smaller \ion{Mg}{2} FWHM for larger $\Gamma$, 
(4) (panel $k$) larger \ion{Mg}{2} FWHM for smaller \fe2($O1$)/\ion{Mg}{2}, 
(5) (panel $m$) larger black hole mass $M_{BH}$\footnote{$M_{BH}$ is 
approximated as (luminosity)$^{0.5}$(\ion{Mg}{2} FWHM)$^2$ \citep{k00}. We 
derived $M_{BH}$ from $M_{BH}/M_{\sun} = 3.37 
(\lambda L_{3000}/10^{37}W)^{0.47}$(\ion{Mg}{2} FWHM/km s$^{-1}$)$^2$ 
\citep{mj}, where $\lambda L_{3000}$ is the luminosity at 3000 \AA.} for 
smaller $\Gamma$, 
(6) (panel $o$) larger $M_{BH}$ for smaller \fe2($O1$)/\ion{Mg}{2}, 
(7) (panel $q$) larger [\ion{O}{3}]/H$\beta$ for larger \ion{Mg}{2} FWHM, 
and (8) (panel $x$) larger \ion{Fe}{2}($O1$)/\ion{Mg}{2} for larger 
\ion{Fe}{2}($O1$)/\ion{Fe}{2}($U1$). We note that the correlation of smaller 
\mg2\ FWHM for larger $\Gamma$ in panel $i$ is consistent with the 
correlation of smaller H$\beta$ FWHM for larger $\Gamma$ \citep{wan, law, 
l97a}.

The fact that six of eight correlations with a confidence level of 99\%
or higher are related to FWHM or $M_{BH}$ ($\propto$ FWHM$^2$) may imply
that $M_{BH}$ is a fundamental quantity that controls $\Gamma$ or SED of
the incident continuum, which in turn controls the \fe2\ emission. Panel 
$m$ indicates that AGNs with smaller $M_{BH}$ have larger excess of soft X-rays, 
resulting in stronger \ion{Fe}{2} emission, especially in the optical as 
shown in panel $n$ and $o$. This may agree with the result obtained for I Zw 1, 
because NLSy1s are a class of AGNs that harbor a low-mass BH at the center 
radiating near the Eddington limit (e.g., Boroson 2002).  
It is also interesting that the optical \fe2\ emission correlates better with
other parameters than UV \fe2\ emission and that \ion{Fe}{2}($O1$)/\ion{Fe}{2}($U1$) 
tightly correlates with \ion{Fe}{2}($O1$)/\ion{Mg}{2}, but not with 
\ion{Fe}{2}($U1$)/\ion{Mg}{2}. These may indicate that UV and optical \ion{Fe}{2} 
emission are generated in different regions of BELR clouds. 
The correlation of larger \mg2\ FWHM for brighter $M_B$ is consistent with 
a weak trend of larger FWHM of Balmer lines for higher luminosity found 
by various authors (e.g., Shuder 1984; Wandel \& Yahil 1985; Kaspi et al. 2000). 

Note that the results of linear regression analysis presented here were derived
from the spectra with no correction for the intrinsic extinction. 
To evaluate the effect of intrinsic extinction, we analyzed the spectra of two 
quasars, PG 1114$+$445 and I Zw 1, with correction for the SMC-type
extinction. 
The resulting change of confidence level is only within 4\%, 
except for the result in panel $c$ and $d$ where the confidence level is 
decreased from 85\% to 62\% and increased from 18\% to 97\%, respectively. 
This is attributed to the difference in the Balmer continuum flux of I Zw 1, 
e.g., Bac/H$\beta$ is 23.4 (without intrinsic extinction) or 2.3 
(with intrinsic extinction).

\section{Summary} 

We have presented the spectra of 14 quasars with a wide wavelength coverage from 1000 to 7300 \AA\ in the rest frame. The redshift ranges from $z =$ 0.061 to 0.555 and the luminosity from $M_{B}$ = $-$22.69 to $-$26.32. These spectra are in high quality and resulted from combining {\it HST} spectra with those taken on ground-based telescopes. 
We described the procedure of deriving the template spectrum of \ion{Fe}{2} from the I Zw 1 spectrum, 
covering two wavelength regions of 2200$-$3500 \AA\ and 4200$-$5600 \AA\ where prominent \ion{Fe}{2} 
emission is seen.  Our \fe2\ template spectrum is semi-empirical in the sense that the synthetic spectrum 
calculated with the CLOUDY photoionization code is used to separate the \fe2\ emission from the \mg2 $\lambda$2798 line.

Our procedure of measuring strengths of \fe2\ emission is twofold; (1) subtracting the continuum component by 
fitting models of the power-law and Balmer continua to the continuum windows which are relatively free from 
line emissions, and (2) fitting models of the \fe2\ emission made by the \fe2\ 
template to the continuum-subtracted spectra. From 14 quasars including I Zw 1, we obtained the \fe2\ fluxes 
in five wavelength bands from the UV to optical, the total flux of Balmer continuum, 
the flux and FWHM of \mg2 $\lambda$2798, H$\alpha$, and other emission lines. 

Regression analysis was performed to derive a linear relation between two variables ($Y = A + BX$), 
and eight correlations with a confidence level of 99\% or higher were 
found. These are the correlations of (1) larger \ion{Mg}{2} FWHM for
larger H$\alpha$ FWHM, (2) larger $\Gamma$ for fainter $M_B$, (3)
smaller \ion{Mg}{2} FWHM for larger $\Gamma$, (4) larger \ion{Mg}{2}
FWHM for smaller \fe2($O1$)/\ion{Mg}{2}, (5) larger $M_{BH}$ for smaller
$\Gamma$, (6) larger $M_{BH}$ for smaller \fe2($O1$)/\ion{Mg}{2}, 
(7) larger [\ion{O}{3}]/H$\beta$ for larger \ion{Mg}{2} FWHM, 
and (8) larger \ion{Fe}{2}($O1$)/\ion{Mg}{2} for larger 
\ion{Fe}{2}($O1$)/\ion{Fe}{2}($U1$). Six of these eight correlations are 
related to FWHM or $M_{BH}$ ($\propto$ FWHM$^2$). This fact may imply that 
$M_{BH}$ is a fundamental quantity that controls $\Gamma$ or equivalently SED 
of the incident continuum that eventually controls \fe2\ emission.

\begin{figure}
\epsscale{0.9}
\plotone{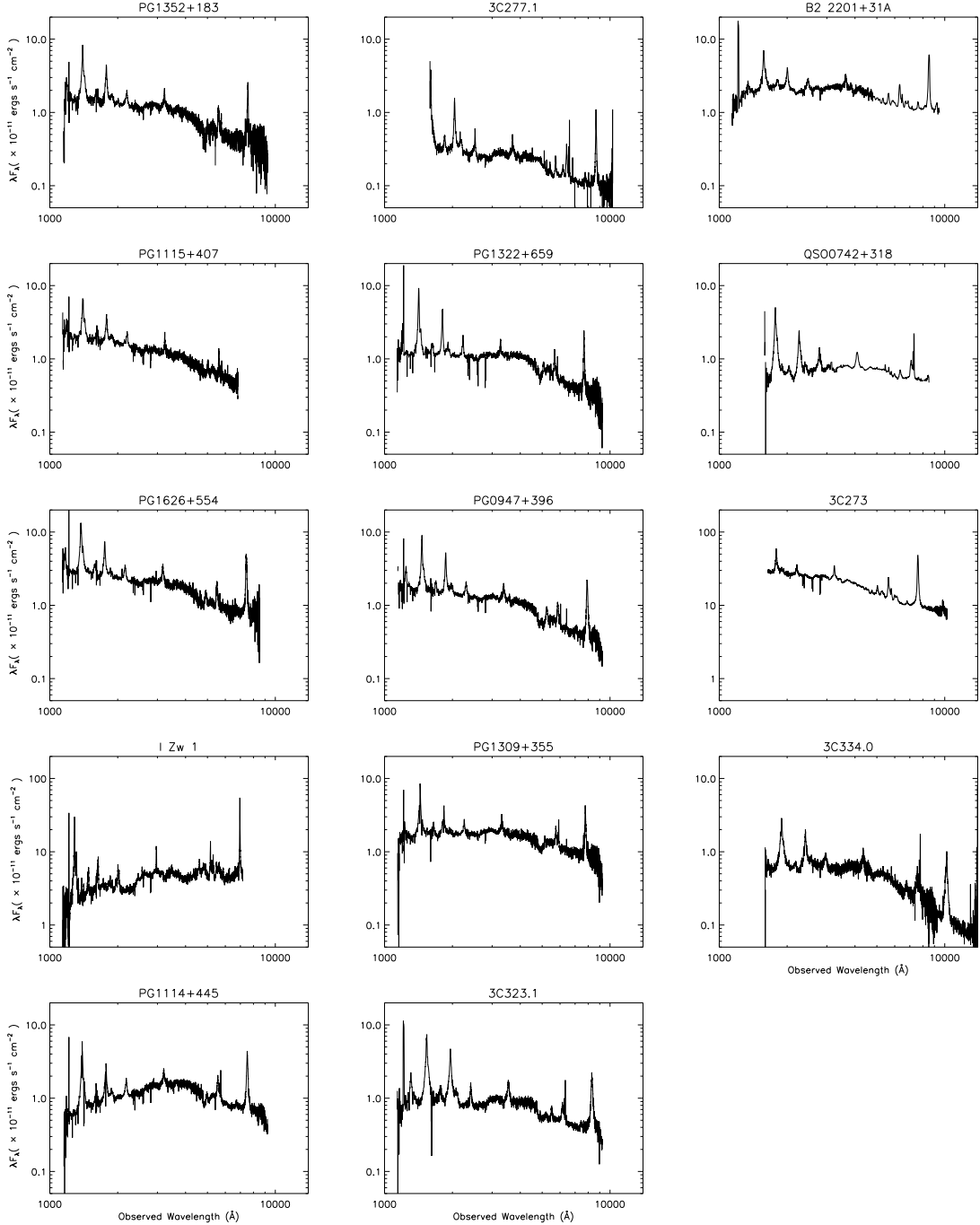}
\caption{Combined UV, optical, and near-infrared spectra in the observer's 
 frame, with no correction for the Galactic extinction.  $\lambda F_{\lambda}$ 
 in $10^{-11}$ \ferg\ is plotted against $\lambda$ in \AA. 
 The spectra are sorted by the $B$ magnitude from PG 1352+183 of the lowest 
 luminosity (top left), then PG 1115+407 of the second lowest 
 luminosity (second-top left), to 3C 334.0 of the highest luminosity (bottom 
 right). The original optical spectra of QSO 0742+318 published in 
 \citet{wnw} and B2 2201+31A in \citet{c97} have been corrected for the 
 Galactic extinction, therefore they are reddened by the same amount of the 
 Galactic extinction as these authors applied.  
 All the spectra will be available from the web site:
http://www.ioa.s.u-tokyo.ac.jp/$^\sim$kkawara/quasars/index.html.\label{ObservedSpectra}}
\end{figure}

\begin{figure}
\epsscale{0.6}
\plotone{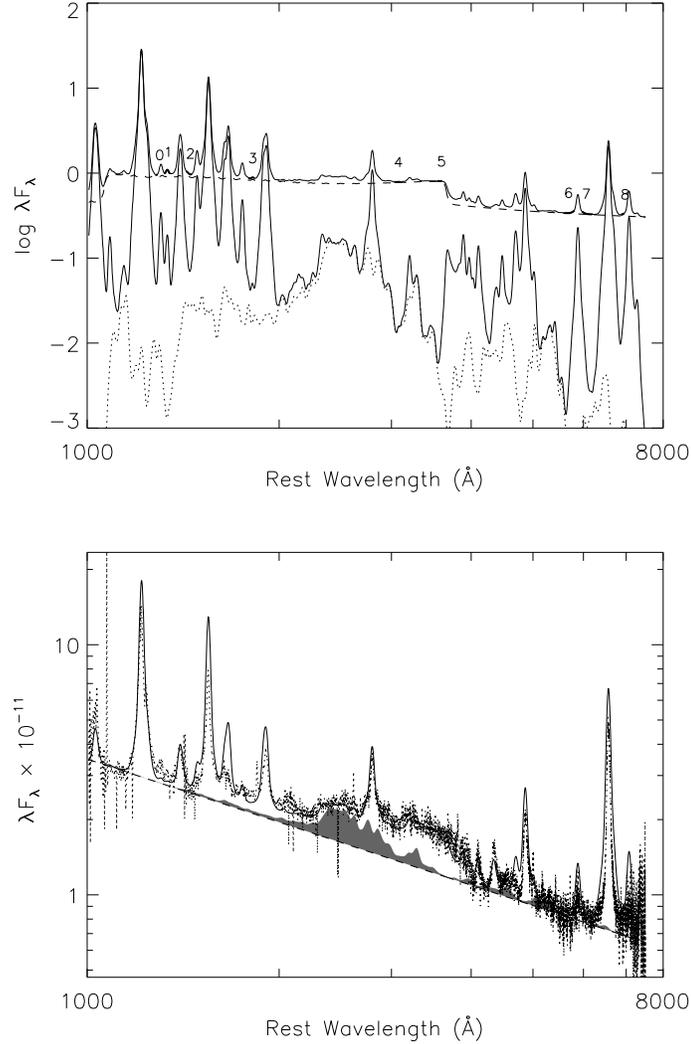}
\caption{\label{modelspec} 
 Upper panel: Model-A spectrum simulated for a single cloud with $\alpha(UV)$ = $-$0.2, $\Gamma$ = 2.8, $\alpha(OX)$ = $-$1.4, $T_{cut}$ = 1.5 $\times 10^5$ K, and $kT_{IR}$ = 0.136 eV, together with gas density of $N_H = 10^{10}$ cm$^{-3}$, ionizing parameter of $U = 10^{-1}$, microturbulence of $V_{turb}$ = 5 \kms, and the solar abundance. 
The spectrum by solid line on the top is the total flux of the model spectrum, while the spectrum by dashed line is the Balmer and Paschen continua. The spectrum by solid line in the middle is the total flux of all line emissions including \ion{Fe}{2}, and the spectrum by dotted line is the contribution from \ion{Fe}{2} emission alone. The numbers denotes the location of individual continuum windows (CWs; see Table \ref{cws}). Lower panel: Model-B spectrum ({\it solid line}) compared with the observed spectrum of PG 1626+554 ({\it dotted line}). This model spectrum is made by applying the SMC-like intrinsic extinction of $E_{B-V}$ = 0.13 to the Model-A spectrum in the upper panel and adding a power-law continuum without extinction. Extinction, power-law index, and a covering factor of 0.5 are found as a result of adjustment to reproduce the observed spectrum of PG 1626+554. The power-law continuum is plotted by dashed line, and \ion{Fe}{2} emission is shown by shaded areas. }
\end{figure}

\begin{figure}
\epsscale{0.7}
\plotone{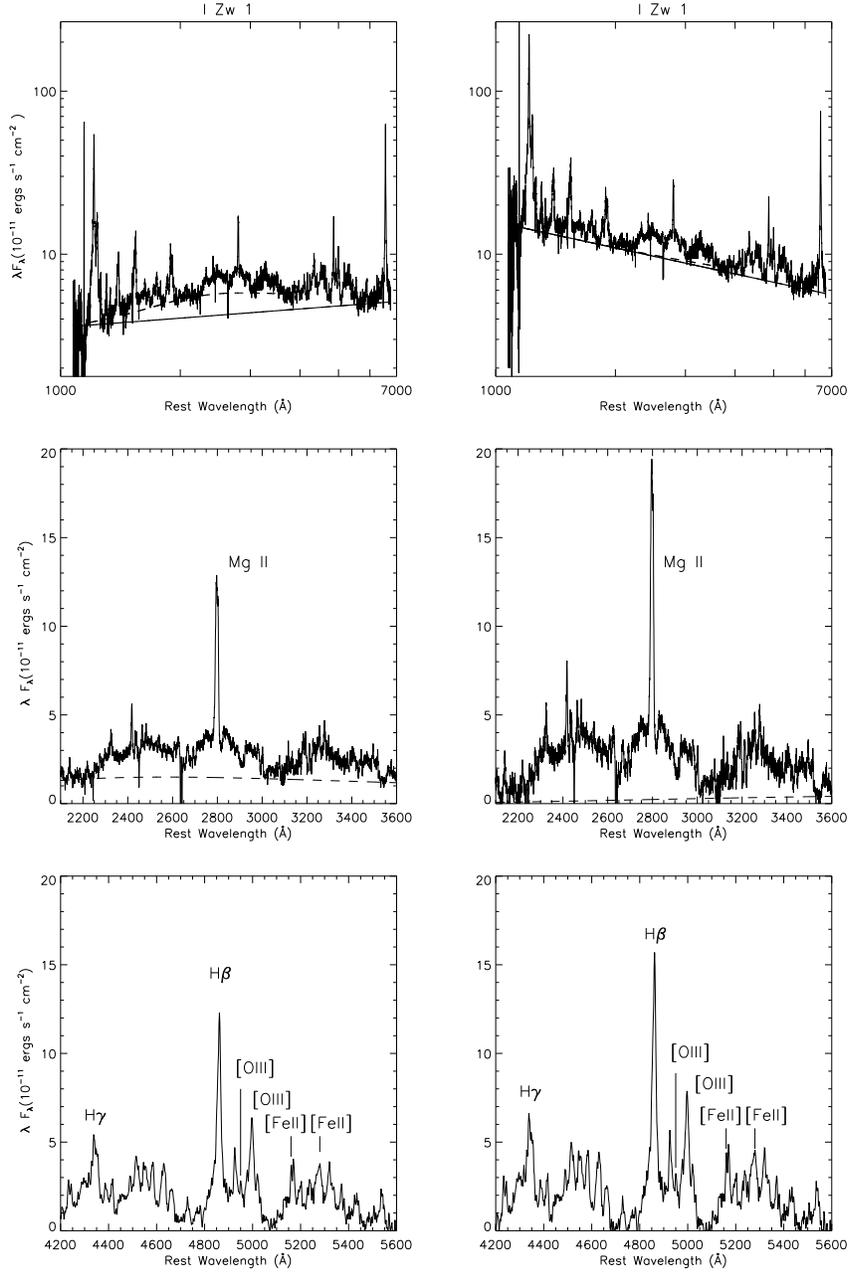}
\caption{\label{zw1spec} 
 Spectrum of I Zw 1, where the observed flux in $10^{-11}$ ergs s$^{-1}$ cm$^{-2}$ 
 is plotted against rest wavelength in \AA.  The spectra in the left panels 
 are with no correction for the intrinsic extinction, while those in the right 
 panels are corrected for the SMC-like intrinsic extinction inferred from 
 the \ion{O}{1} line ratio. The correction for Galactic extinction has been 
 applied to all the spectra. 
 The top panels show the spectra to which the model continua are fitted. 
 The best-fit continua are shown by the solid line for the power-law continuum 
 and by the dashed line for the Balmer continuum.
 The middle and bottom panels show the power-law subtracted spectra in the UV 
 and optical regions, respectively. The dashed line in the middle panels 
 represent the best-fit Balmer continuum. The spectra in the middle and bottom 
 panels are dominated by \ion{Fe}{2} emission.}
\end{figure}

\begin{figure}
\plotone{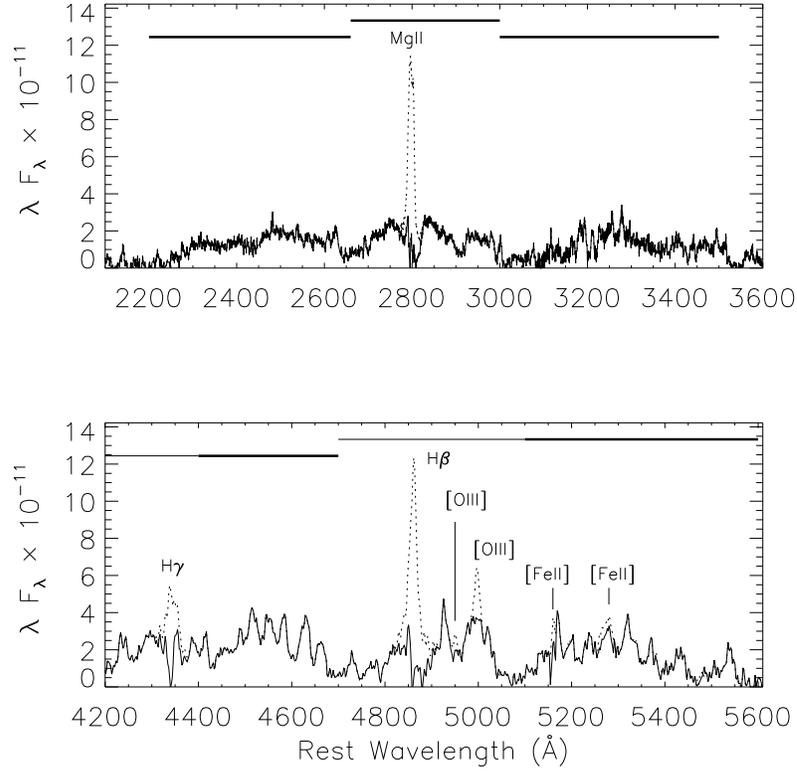}
\caption{\label{zw1fe2} 
Spectrum of I Zw 1 in the UV (upper panel) and optical (lower panel) regions, 
where the observed flux in $10^{-11}$ ergs s$^{-1}$ cm$^{-2}$ is plotted against 
rest wavelength in \AA, after subtracting the power-law and Balmer continua 
and correcting for the Galactic extinction only.  
The solid lines show the \ion{Fe}{2} spectra, while the dotted lines show the 
contribution of other emission lines. Thick horizontal bars indicate five 
regions where template fitting was performed and \ion{Fe}{2} strengths were 
measured, while thin horizontal bars indicate two regions where 
non-\ion{Fe}{2} emission lines were measured after the \ion{Fe}{2} spectrum was subtracted. 
}
\end{figure}

\begin{figure}
\plotone{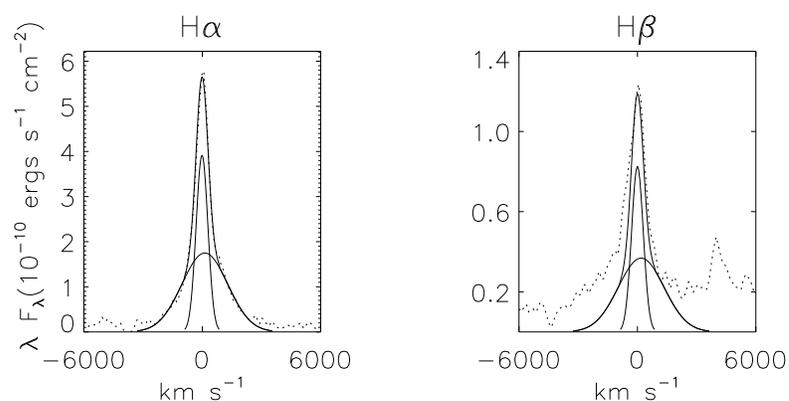}
\caption{\label{zw1ha} 
Emission line profiles for H$\alpha$ (left panel) and H$\beta$ (right panel) 
in the continuum-subtracted spectrum of I Zw 1, where the observed 
flux in $10^{-10}$ ergs s$^{-1}$ cm$^{-2}$ is plotted against relative velocity in \kms. 
The dotted lines represent the observed profiles.  Two Gaussian components 
with FWHMs of 690 and 2700 km s$^{-1}$ and the peak height ratio of 1:0.4 are 
best fitted to the H$\alpha$ profile.  The two components are indicated by the 
thin lines, and their sum by the thick line. Note that the 690 km s$^{-1}$ 
component is blueshifted relative to the 2700 km s$^{-1}$ component by 
150 \kms. The two-component H$\alpha$ profile is used as the H$\alpha$ 
template for our analysis.} 
\end{figure}

\begin{figure}
\epsscale{0.6}
\plotone{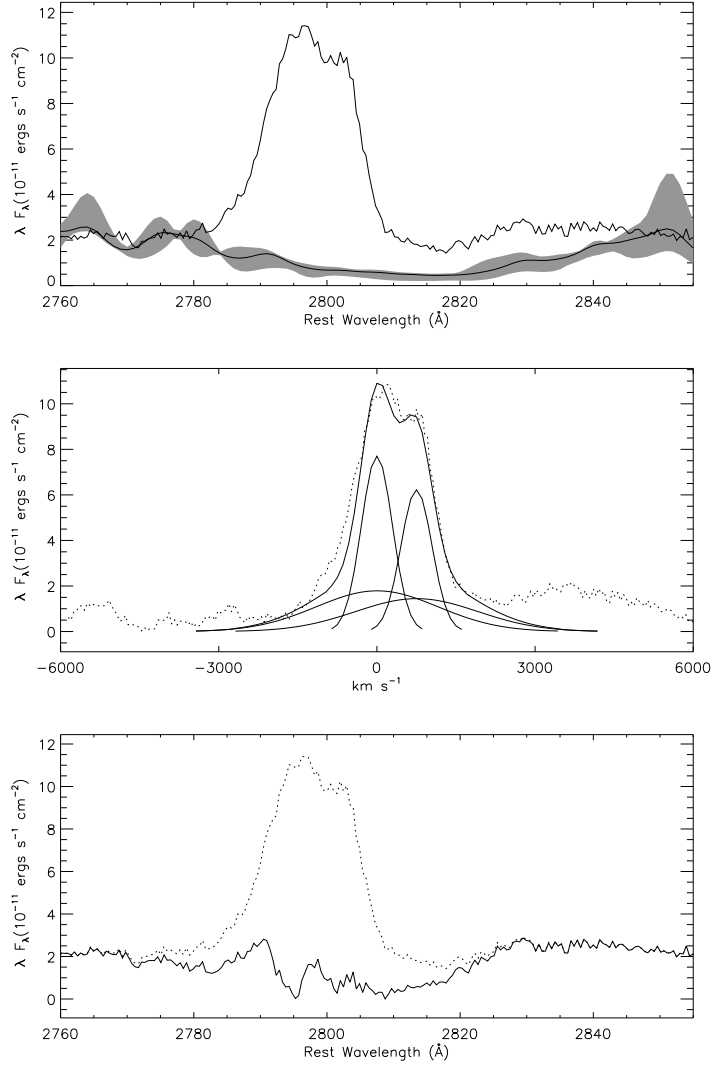}
\caption{\label{zw1mg} 
Top panel shows the Model-A spectrum ({\it thick solid line}) of \ion{Fe}{2} 
emission overplotted on the I Zw 1 spectrum ({\it thin solid line}) around 
\ion{Mg}{2} $\lambda$2798. The model spectrum was convolved with a 690 \kms
FWHM profile. The shaded area is a 
range allowed by synthetic \ion{Fe}{2} spectra in the parameter space used to 
study the continuum windows in section \ref{Continuum Windows}. 
The middle panel shows the \ion{Mg}{2} $\lambda$2798 fit ({\it thick solid line}) to 
the spectrum ({\it dotted line}) where the model spectrum of \ion{Fe}{2} has been 
subtracted. The \ion{Mg}{2} $\lambda$2798 fit consists of doublet lines at 
2795.5 and 2802.7 {\AA} and each line has two Gaussian components ({\it thin solid 
lines}) with FWHMs of 690 and 2700 km s$^{-1}$ as in the case of the H$\alpha$ 
template. The intensity ratio of the doublet lines is 1.2:1 and the peak height ratio of two Gaussian components is 1:0.23 
with no velocity difference between them.  The bottom panel shows the spectrum 
({\it thin solid line}) of \ion{Fe}{2} emission overplotted on the I Zw 1 spectrum
({\it dotted line}). This \ion{Fe}{2} spectrum was obtained by 
subtracting the \ion{Mg}{2} $\lambda$2798 fit from the observed spectrum.} 
\end{figure}

\begin{figure}
\plotone{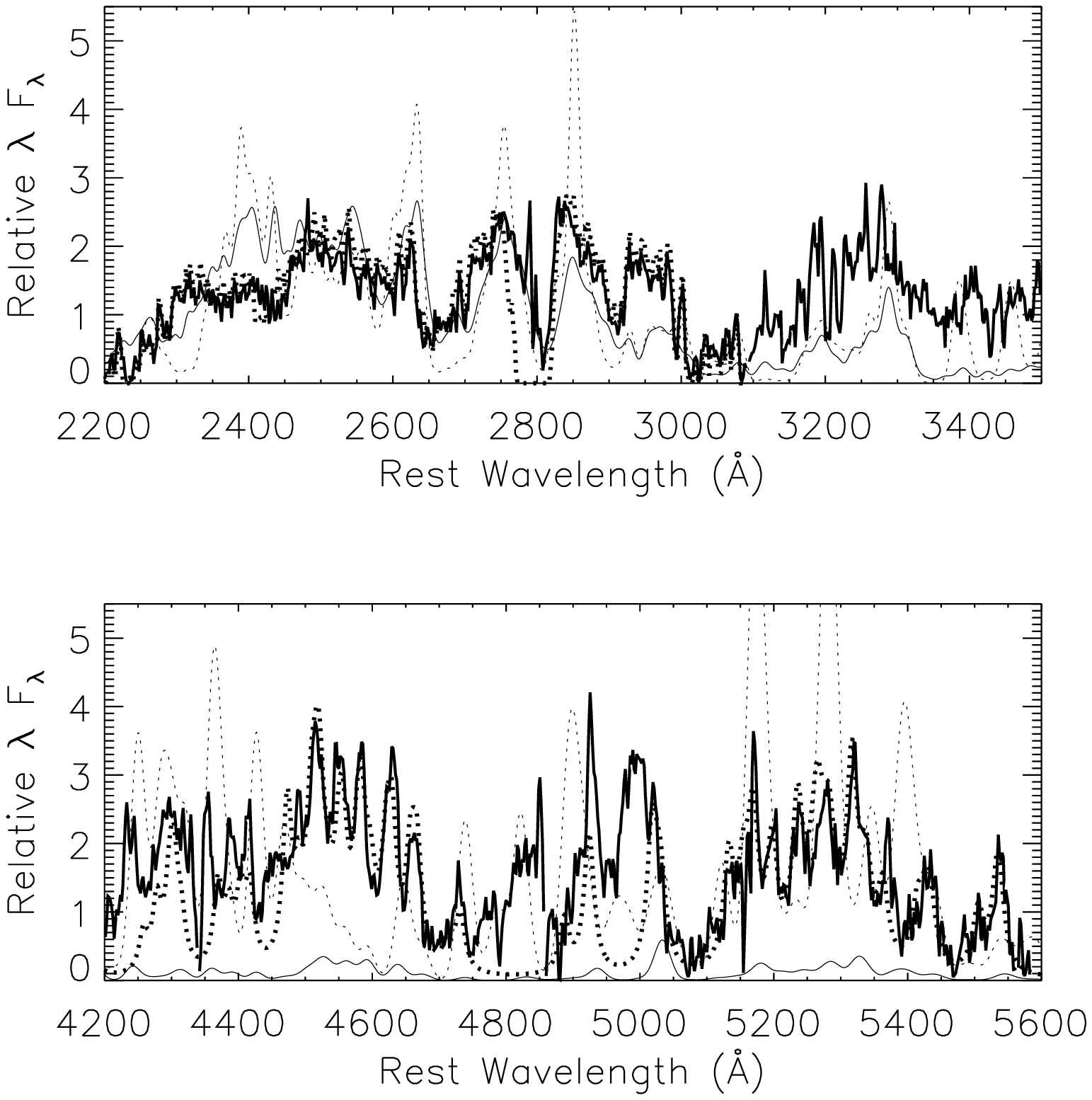}
\caption{\label{zw1comp} 
Our \ion{Fe}{2} template spectrum ({\it thick line}) is compared with 
the UV template by \citet{vw} ({\it thick dots}) in the upper panel 
and the optical template by \citet{vjv} ({\it thick dots})
in the lower panel. Two synthetic spectra on model-A  ({\it thin line}) 
and a LDC model  ({\it thin dots}) in the 
framework of photoionization are overplotted. The spectra are scaled in flux,
in such a way that the fluxes integrated between 2200 and 3100 {\AA} matches 
with our template spectrum. 
The template by \citet{vjv} are given in relative units, and so it was 
scaled in flux to match with our template spectrum 
in 4400$-$4700 \AA\ and 5100$-$5600 \AA. The synthetic spectra were 
convolved with a 1660 km s$^{-1}$ FWHM Gaussian profile. 
Our \ion{Fe}{2} spectrum is available from the web site: 
            http://www.ioa.s.u-tokyo.ac.jp/$^\sim$kkawara/quasars/index.html.}
\end{figure}

\begin{figure}
\plotone{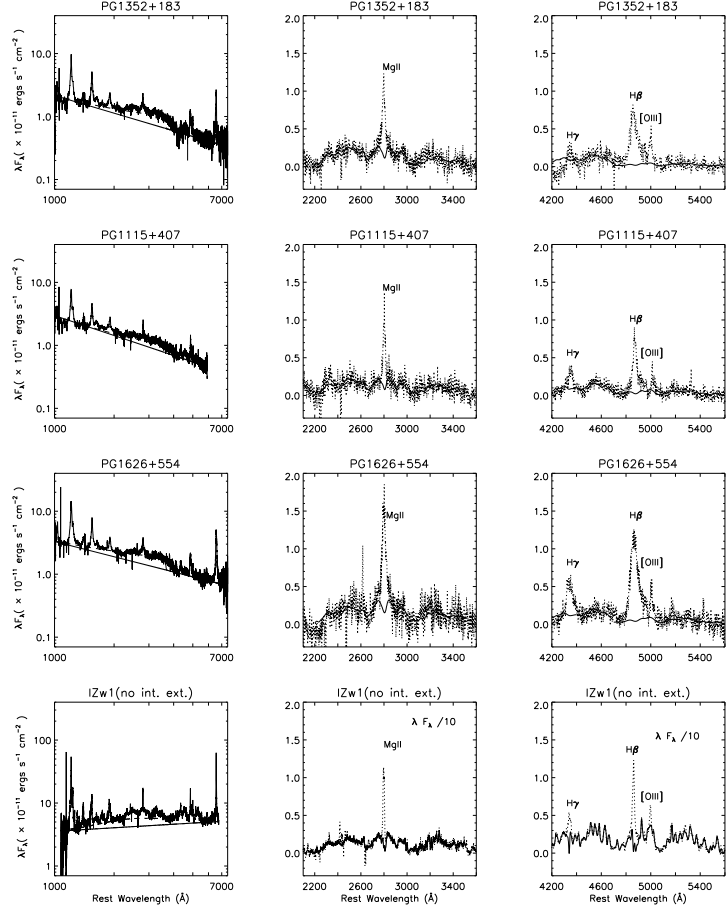} 
\caption{\label{qso1}
The best-fit models of 14 low-redshift quasars overplotted on the original 
spectra, where the observed flux in $10^{-11}$ ergs s$^{-1}$ cm$^{-2}$ is 
plotted against rest wavelength in \AA. 
The panels from the top row to the bottom are arranged in order of decreasing 
luminosity, namely, PG 1352+183 of the lowest luminosity shown in three panels
in the top row.  For each quasar, the best-fit continuum model is overplotted 
on the original quasar spectrum ({\it left panel}), and the best-fit 
\ion{Fe}{2} emission models are plotted on the continuum-subtracted spectra 
in the UV ({\it middle}) and optical ({\it right}). 
Two fitting cases are shown for I Zw 1 and PG 1114+445; one is with no 
intrinsic extinction, and the other is with SMC-like intrinsic 
extinction of $E_{B-V} = 0.09$. In the left panels, the solid and dashed lines 
represent the power-law and Balmer continua, respectively.  In the middle and 
left panels, the thick and thin lines represent the best-fit \ion{Fe}{2} 
models.  Note that two kinds of lines are used here to distinguish the 
\ion{Fe}{2} bands by eyes.} 
\end{figure}

\begin{figure}
\plotone{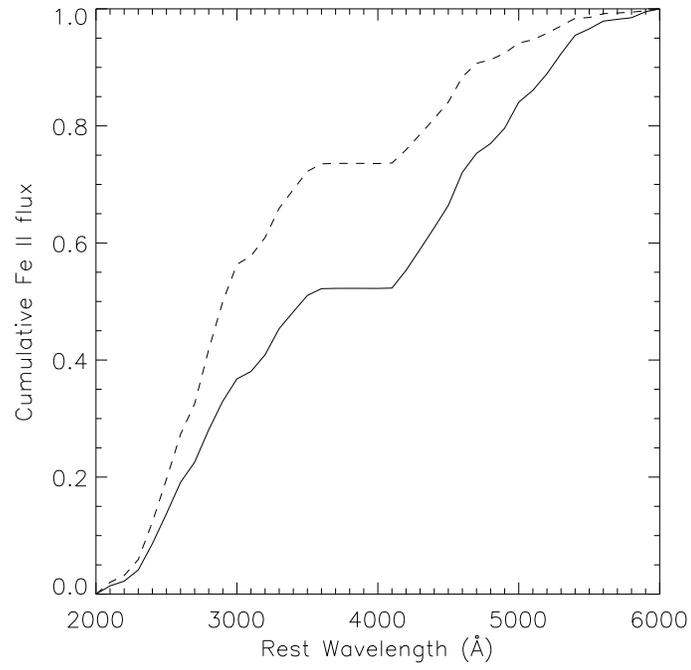}
\caption{\label{cumulative_fe2}
Cumulative \ion{Fe}{2} flux for I Zw 1 ({\it solid line}) and PG 1626+554 
({\it dashed line}). 
The total \ion{Fe}{2} strength integrated 
from 2000 to 6000 \AA\ is normalized to unity.} 
\end{figure}

\begin{figure}
\epsscale{0.9}
\plotone{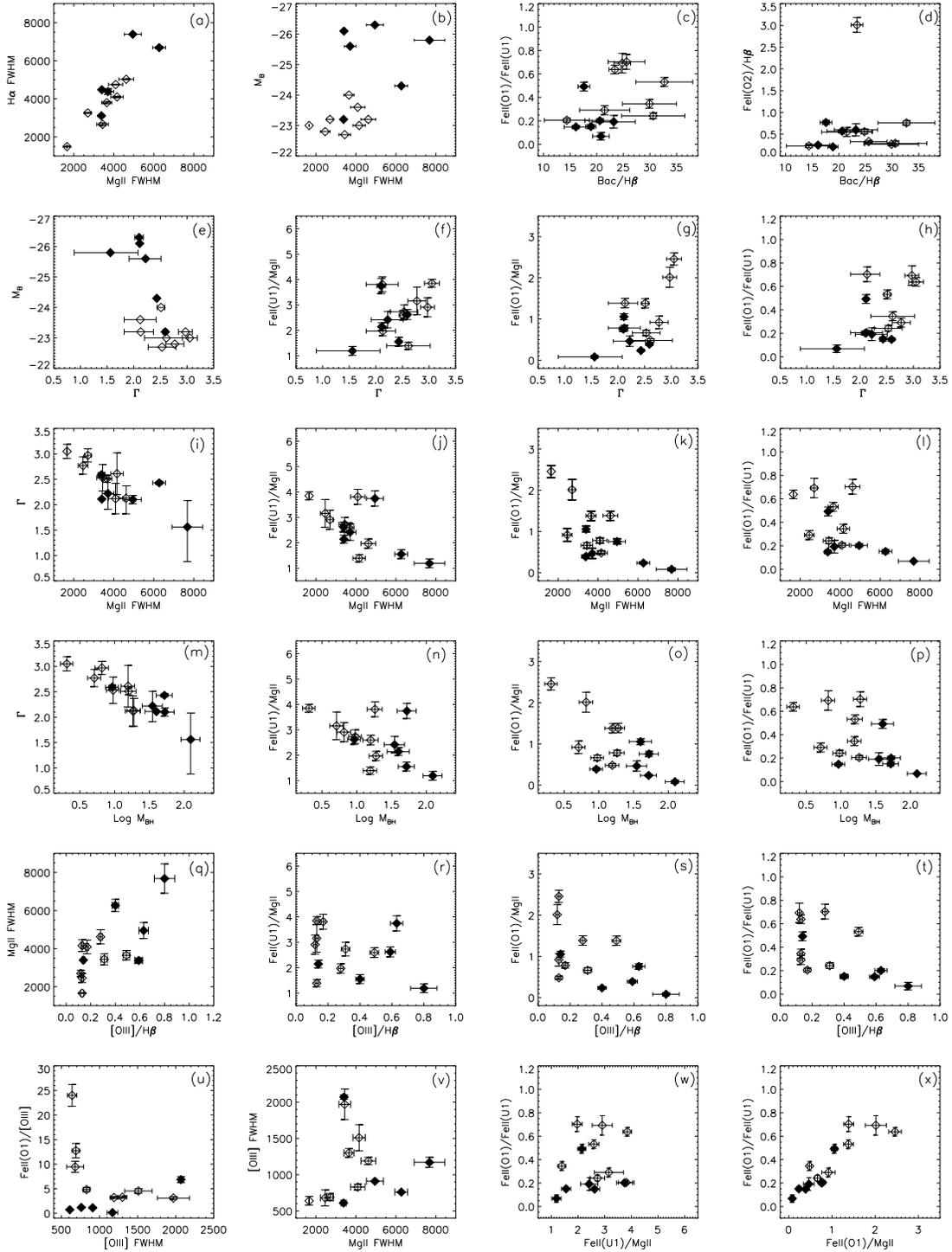} 
\caption{\label{correlations}
Correlations related to \ion{Fe}{2} emission in 14 quasars in our sample. Open and filled symbols are for radio-quiet and radio-loud quasars, respectively. FWHM is in km s$^{-1}$\ and the definition of the X-ray photon index $\Gamma$ is given in the notes of Table \ref{multi}. The black hole mass $M_{BH}$ is in units of $10^7 M_{\sun}$ and derived from $M_{BH}/M_{\sun} = 3.37 (\lambda L_{3000}/10^{37}W)^{0.47}$ (\ion{Mg}{2} FWHM/km s$^{-1}$)$^2$ \citep{mj}. \ion{Fe}{2}($U1$), \ion{Fe}{2}($O1$), and \ion{Fe}{2}(O2) are the fluxes in the wavelength bands of $U1$ (2200$-$2660 \AA), $O1$ (4400$-$4700 \AA), and $O2$ (5100$-$5600 \AA), respectively.} 
\end{figure}

Figure \ref{ObservedSpectra} and the \ion{Fe}{2} template are available from 
the web site:  \newline {\small http://www.ioa.s.u-tokyo.ac.jp/$^\sim$kkawara/quasars/index.html}. \newline 

\acknowledgements
We thank M. Vestergaard for kindly providing their UV template spectrum 
of iron emission in the electric form, G.J. Ferland for his support for 
our use of the CLOUDY photoionization code, and the anonymous referee 
for useful and constructive comments. This work has been supported in part by 
Grant-inAid for Scientific research (3730, 12440052, 12440052) from JPSP and 
Center-of-Excellence (COE) research (07CE2002) of the Ministry of Education, 
Science, and Culture of Japan. Y.T. thanks the Hayakawa 
funds in Astronomical Society of Japan for providing travel expenses to visit to the KPNO.


\clearpage

\begin{deluxetable}{lccccc}
\tablecaption{Quasar Sample \label{sample}}
\tablewidth{0pt}
\tablehead{
\colhead{Quasar} & \colhead{$\alpha$ (J2000.0)} & \colhead{$\delta$ (J2000.0)} & \colhead{Redshift} & \colhead{$M_{B}$\tablenotemark{a}} & \colhead{$E_{B-V}$\tablenotemark{b}} 
}
\startdata
I Zw 1        & 00 53 34.9 & +12 41 36    & 0.061    & $-$22.97 &  0.06  \\
QSO 0742+318   & 07 45 41.7 & +31 42 56    & 0.462    & $-$25.76 &  0.06  \\
PG 0947+396    & 09 50 48.4 & +39 26 51    & 0.206    & $-$23.56 &  0.01  \\
PG 1114+445    & 11 17 06.3 & +44 13 34    & 0.144    & $-$23.22 &  0.01  \\
PG 1115+407    & 11 18 30.4 & +40 25 55    & 0.154    & $-$22.83 &  0.02  \\
3C 273         & 12 29 06.6 & +02 03 08    & 0.158    & $-$26.13 &  0.02  \\
3C 277.1       & 12 52 26.4 & +56 34 19    & 0.320    & $-$23.22 &  0.01  \\
PG 1309+355    & 13 12 17.7 & +35 15 23    & 0.184    & $-$24.02 &  0.01  \\
PG 1322+659    & 13 23 49.6 & +65 41 48    & 0.168    & $-$23.24 &  0.02  \\ 
PG 1352+183    & 13 54 35.6 & +18 05 18    & 0.152    & $-$22.69 &  0.02  \\
3C 323.1       & 15 47 43.6 & +20 52 16    & 0.266    & $-$24.33 &  0.05  \\
3C 334.0       & 16 20 21.8 & +17 36 23    & 0.555    & $-$26.32 &  0.05  \\
PG 1626+554    & 16 27 56.2 & +55 22 32    & 0.132    & $-$22.97 &  0.01  \\
B2 2201+31A   & 22 03 14.9 & +31 45 38    & 0.298    & $-$25.60 &  0.12
 \\ 
\enddata
\tablenotetext{a}{Based on $B$ magnitudes derived from integrating the
 flux density of our spectra in the $B$-band.}
\tablenotetext{b}{Galactic extinction $E_{B-V}$ taken from \citet{sfd}.}
\end{deluxetable}

\begin{deluxetable}{lccrc}
\tabletypesize{\small}
\tablecaption{{\it HST} Observations \label{hst}}
\tablewidth{0pt}
\tablehead{
\colhead{Quasars} & \colhead{$\lambda$ range} & \colhead{Grating} &
 \colhead{$t_{int}^a$} & \colhead{Date} \\
\colhead{}        & \colhead{($\mbox{\AA}$)}  & \colhead{}        &
 \colhead{(sec)}       & \colhead{} 
}
\startdata
I Zw 1      & 1140$-$1606 & FOS G130H   & 29880.0  & 1994 Feb 13 \\
            & 1590$-$2312 & FOS G190H   & 6120.0   & 1994 Sep 14 \\
            & 2222$-$3277 & FOS G270H   & 2160.0  & 1994 Sep 14 \\
QSO 0742+318 & 1590$-$2312 & FOS G190H   & 4483.9  & 1994 Mar 14 \\
            & 2222$-$3277 & FOS G270H   & 1598.5  & 1994 Mar 14 \\
PG 0947+396  & 1140$-$1606 & FOS G130H   & 4007.0  & 1997 May 06 \\
            & 1590$-$2312 & FOS G190H   & 1260.0  & 1997 May 06 \\
            & 2222$-$3277 & FOS G270H   & 450.0   & 1997 May 06 \\
            & 3235$-$4781 & FOS G400H   & 210.0   & 1997 May 06 \\
PG 1114+445  & 1140$-$1606 & FOS G130H   & 9269.8  & 1996 Mar 13\\
            & 1590$-$2312 & FOS G190H   & 1560.0  & 1996 Mar 13\\
            & 2222$-$3277 & FOS G270H   & 160.0   & 1996 Mar 13\\
            & 3235$-$4781 & FOS G400H   & 120.0   & 1996 Mar 13\\
PG 1115+407  & 1140$-$1606 & FOS G130H   & 3749.9  & 1996 Mar 19 \\
            & 1590$-$2312 & FOS G190H   & 760.0   & 1996 Mar 19 \\
            & 2222$-$3277 & FOS G270H   & 191.0   & 1996 Mar 19 \\
            & 3235$-$4781 & FOS G400H   & 154.0   & 1996 Mar 19 \\
            & 4569$-$6818 & FOS G570H   & 471.0   & 1996 Mar 19 \\
3C 273       & 1140$-$1606 & FOS G130H   & 7600.0  & 1991 Feb 17 \\
            & 1590$-$2312 & FOS G190H   & 5414.4  & 1991 Feb 14, 15\\
            & 2222$-$3277 & FOS G270H   & 5414.4  & 1991 Feb 15 \\
            & 2900$-$5700  & STIS G430L & 974.0   & 2002 Feb 01\\
            & 5240$-$10270 & STIS G750L & 776.0   & 2002 Feb 01\\
3C 277.1     & 1590$-$2312 & FOS G190H   & 1986.0  & 1993 Jul 01\\
            & 2222$-$3277 & FOS G270H   & 1122.0  & 1993 Jul 01\\
            & 3235$-$4781 & FOS G400H   & 864.0   & 1993 Jul 01\\
            & 2900$-$5700   & STIS G430L & 2800.0  & 1999 Oct 12\\
            & 5240$-$10270  & STIS G750L & 2340.0  & 1999 Oct 11\\
PG 1309+355  & 1140$-$1606 & FOS G130H   & 4580.0  & 1997 May 20 \\
            & 1590$-$2312 & FOS G190H   & 799.0   & 1997 May 20 \\
            & 2222$-$3277 & FOS G270H   & 306.0   & 1997 May 20 \\
            & 3235$-$4781 & FOS G400H   & 112.0   & 1997 May 20\\
 PG 1322+659 & 1140$-$1606 & FOS G130H   & 22147.6 & 1997 Jan 19\\
            & 1590$-$2312 & FOS G190H   & 2549.9  & 1997 Jan 19\\
            & 2222$-$3277 & FOS G270H   & 416.0   & 1997 Jan 19\\
            & 3235$-$4781 & FOS G400H   & 334.0   & 1997 Jan 19\\
 PG 1352+183 & 1140$-$1606 & FOS G130H   & 2170.0 & 1996 Mar 26\\
            & 1590$-$2312 & FOS G190H   & 702.0  & 1996 Mar 26\\
            & 2222$-$3277 & FOS G270H   & 200.0  & 1996 Mar 26\\
            & 3235$-$4781 & FOS G400H   & 156.0  & 1996 Mar 26\\
 3C 323.1    & 1140$-$1606 & FOS G130H   & 2310   & 1993 Jul 01 \\
            & 1590$-$2312 & FOS G190H   & 384.0  & 1993 Jul 01 \\
            & 2222$-$3277 & FOS G270H   & 225.0  & 1993 Jul 01 \\
            & 3235$-$4781 & FOS G400H   & 156.0  & 1993 Jul 01 \\
 3C 334.0    & 1590$-$2312 & FOS G190H   & 648.0  & 1993 Jul 01 \\
            & 2222$-$3277 & FOS G270H   & 345.0  & 1993 Jul 01 \\
            & 3235$-$4781 & FOS G400H   & 252.0  & 1993 Jul 01 \\
 PG 1626+554 & 1140$-$1606 & FOS G130H   & 3192.9 & 1996 Nov 19 \\
            & 1590$-$2312 & FOS G190H   & 654.0  & 1996 Nov 19 \\
            & 2222$-$3277 & FOS G270H   & 120.0  & 1996 Nov 19 \\
            & 3235$-$4781 & FOS G400H   & 92.0   & 1996 Nov 19 \\
            & 4569$-$6818 & FOS G570H   & 255.0  & 1996 Nov 19 \\
            & 6270$-$8500 & FOS G780H   & 825.0  & 1996 Nov 19 \\
B2 2201+31A & 1140$-$1606 & FOS G130H   & 2790.0 & 1993 Jul 01\\
            & 1590$-$2312 & FOS G190H   & 1104.0 & 1993 Jul 01\\
            & 2222$-$3277 & FOS G270H   & 222.0  & 1993 Jul 01\\
            & 3235$-$4781 & FOS G400H   & 132.0  & 1993 Jul 01\\
\enddata
\tablenotetext{a}{Integration times on target.}
\end{deluxetable}

\begin{deluxetable}{llllrcc}
\tabletypesize{\scriptsize}
\tablecaption{Ground-Based Observations \label{ground}}
\tablewidth{0pt}
\tablehead{
\colhead{Quasars} & \colhead{$\lambda$ range} & \colhead{$\Delta\lambda$} & \colhead{Spectrograph\tablenotemark{a}} & \colhead{$t_{int}$\tablenotemark{b}} &
 \colhead{Date\tablenotemark{c}} & \colhead{References\tablenotemark{d}}
 \\
\colhead{}       & \colhead{($\mbox{\AA}$)}  & \colhead{(\AA)} &
 \colhead{} & \colhead{(sec)} & \colhead{} & \colhead{} 
}
\startdata
I Zw 1      & 3183$-$4073  & 1.2 & KPNO 2.1m Goldcam   &          & 1995 Sep 22 
                                                & \citet{l97b}\tablenotemark{e} \\
            & 4008$-$7172  & 3.1 & INT 2.5m IDS        & 1220     & 1987 Aug 15 & ING archive\\
QSO 0742+318 & 2519$-$8510  & 2.47 & McDonald 2.7m UVITS &         &
 & \citet{wnw} \\
PG 0947+396  & 4330$-$9225  & 2.47 & KPNO 2.1m Goldcam  & 1680      & 2001 May 2,3 & This work \\
PG 1114+445  & 4330$-$9226  & 2.47 & KPNO 2.1m Goldcam  &  660      & 2001 May 2,3 & This work \\
PG 1309+355  & 4331$-$9226  & 2.47 & KPNO 2.1m Goldcam  &  420      & 2001 May 2,3 & This work \\
PG 1322+659  & 4331$-$9226  & 2.47 & KPNO 2.1m Goldcam  &  540      & 2001 May 2,3 & This work \\
PG 1352+183  & 4329$-$9228  & 2.47 & KPNO 2.1m Goldcam  &  540      & 2001 May 2,3 & This work\\
3C 323.1     & 4331$-$9228  & 2.47 & KPNO 2.1m Goldcam  & 1620      & 2001 May 2,3 & This work \\
3C 334.0     & 4334$-$9225  & 2.47 & KPNO 2.1m Goldcam  & 1080      & 2001 May 2   & This work \\
            & 8700$-$13900 & 5.83 & Subaru 8.2m CISCO  &  600      & 2001 May 9   & This work\\
B2 2201+31A & 4472$-$9488  & 2.47 &  KPNO 2.1m Goldcam & 5400      &
 1994 Sep 28  & \citet{c97} \\
\enddata
\tablenotetext{a}{IDS (Intermediate Dispersion Spectrograph); UVITS (UVITS spectrograph); CISCO (Cooled Infrared Spectrograph and Camera for OH-suppressor)}
\tablenotetext{b}{Integration times on target are given
 where available.}
\tablenotetext{c}{Dates of the observations are given
 where available.}
\tablenotetext{d}{New observations by this work are labeled ``This work''.}
\tablenotetext{e}{This spectrum is available at http://physics.technion.ac.il/$^\sim$laor/IZw1/.}
\end{deluxetable}

\begin{deluxetable}{lcrrccc}
\tablecaption{Multi-Wavelength Properties \label{multi}}
\tablewidth{0pt}
\tablehead{
\colhead{Quasars} & \colhead{$B$} & \colhead{$B-V$} & \colhead{$\alpha$} & \colhead{$L_{60\mu m}/L_V$} &
 \colhead{$\Gamma$} & \colhead{Radio} 
}
\startdata
I Zw 1       &  14.28 &  0.318  & $-$1.19 & 2.12   & 3.05 $\pm$ 0.14 & Q \\
QSO 0742+318 &  16.26 &  0.279  & $-$0.73 &        & 1.56$^{+0.52}_{-0.68}$ & L \\
PG 0947+396  &  16.21 & $-$0.043  & $-$0.06 & 1.53   & 2.12 $\pm$ 0.30 & Q \\
PG 1114+445  &  15.82 &  0.102  & $-$1.04 & 0.89   & 2.13$^{+0.24}_{-0.31}$ & Q \\
PG 1115+407  &  16.25 &  0.021  & 0.05    & $<$1.0 & 2.77 $\pm$ 0.17 & Q \\
3C 273       &  13.01 &  0.108  & $-$0.14 & 0.62   & 2.11 $\pm$ 0.01 & L \\
3C 277.1     &  17.71 &  0.025  & $-$0.24 & 1.43   & 2.59 $\pm$ 0.04 & L \\
PG 1309+355  &  15.66 &  0.206  & $-$0.63 & $<$0.51& 2.51 $\pm$ 0.07 & Q \\
PG 1322+659  &  16.13 &  0.030  & $-$0.20 & 0.60   & 2.97 $\pm$ 0.13 & Q \\
PG 1352+183  &  16.33 &  0.026  & $-$0.07 & 1.58   & 2.53 $\pm$ 0.26 & Q \\
3C 323.1     &  16.19 & $-$0.128  & $-$0.21 & 0.40   & 2.43 $\pm$ 0.03 & L \\
3C 334.0     &  16.30 & $-$0.121  & 0.15    & 0.75   & 2.10 $\pm$ 0.08 & L \\
PG 1626+554  &  15.75 & $-$0.011  & $-$0.20 & 0.34   & 2.61 $\pm$ 0.41 & Q \\
B2 2201+31A  &  15.03 & $-$0.085  & $-$0.09 &        & 2.22$^{+0.29}_{-0.31}$ & L \\
\enddata
\tablecomments{Column 2: $B$ magnitudes derived from our spectra after 
corrected for the Galactic extinction given in 
                         Table \ref{sample}.
               Column 3: Same as column 2 but for $B-V$.
               Column 4: Power-law index, $F_{\nu} \propto
		         \nu^{\alpha}$, derived from our de-reddened
		         continuum spectra. Fitting procedure is described in
		         section \ref{Continuum Subtraction}.      
             Column 5: The 60 $\mu$m luminosity relative to the optical 
              V-band, as defined by 
      $L_{60\mu m}/L_V = \nu_{60\mu m}F_{60\mu m}/\nu_VF_V$. 
    The 60 $\mu$m data are taken from \citet{sa}, \citet{bem}, and \citet{ha}. 
               Column 6: X-ray photon index $\Gamma$ taken from
		\citet{bys} and \citet{yu}. 
                         $\Gamma$ is defined by the power-law fit, i.e., 
                         $P_{E}$(photons s$^{-1}$ keV$^{-1}$) $\propto$ $e^{-N_H\sigma_E}$ 
                         $\times$ $E^{- \Gamma}$, 
                         where $N_H\sigma_E$ is the  Galactic absorption.
               Column 7: Radio properties, namely, ``Q'' for radio-quiet and ``L'' for radio-loud, based on 
                         the 5 or 1.4 GHz flux in the FIRST catalog by \citet{bwh}, \citet{kell}, 
                         and \citet{kue}.}
\end{deluxetable}

\begin{deluxetable}{cccc}
\tablecaption{Emission Line Contributions in Continuum Windows \label{cws}}
\tablewidth{0pt}
\tablehead{
\colhead{Name} & \colhead{$\lambda$ range} &
 \colhead{$F_{\lambda}$(lines)/$F_{\lambda}$(total)} &
 \colhead{Adopted} \\
 \colhead{}    & \colhead{({\AA})} &
 \colhead{(\%)} & \colhead{(\%)}
}
\startdata
CW0 & 1280$-$1290 & 4$-$9  &    \\
CW1 & 1320$-$1350 & 5$-$8  & 7 \\
CW2 & 1430$-$1460 & 6$-$9  &    \\
CW3 & 1790$-$1830 & 2$-$5  & 4  \\
CW4 & 3030$-$3090 & 1$-$2  & 1  \\
CW5 & 3540$-$3600 &  1     & 1  \\
CW6 & 5600$-$5800 & 2$-$3  & 3  \\
CW7 & 5970$-$6200 & 2$-$3  & 3  \\
CW8 & 6870$-$6950 & 7$-$8 &    \\
\enddata
\tablecomments{Fractional contributions from all emission lines to the 
total flux of synthetic spectra.  Column 3 gives the values expressed in 
percentage, based on our single cloud models calculated for 810 grids 
in the parameter space. Column 4 gives the values adopted for our use of 
continuum subtraction.}
\end{deluxetable}

\begin{deluxetable}{llrcllccccccl}
\tabletypesize{\scriptsize}
\rotate
\setlength{\tabcolsep}{0.05in}
\tablecaption{Parameters for Fitting \label{fitparameter}}
\tablewidth{0pt}
\tablehead{
\colhead{} & \multicolumn{2}{c}{Power-law continuum} & \colhead{} &
 \multicolumn{3}{c}{Balmer continuum} & \colhead{} &
 \multicolumn{5}{c}{\ion{Fe}{2} emission} \\
\cline{2-3} \cline{5-7} \cline{9-13}
\colhead{Quasars} & \colhead{$F_{0}$\tablenotemark{a}} &
 \colhead{$\alpha$(UV)} & \colhead{} & \colhead{$T_{e}$ (10$^{4}$K)} &
 \colhead{$\tau_{BE}$} & \colhead{$h_{BE}$\tablenotemark{b}} &
 \colhead{} & \colhead{$b$} & \colhead{$b_{1}$} & \colhead{$b_{2}$} &
 \colhead{$b_{3}$} & \colhead{$b_{4}$} 
}
\startdata
\multicolumn{13}{c}{With no intrinsic extinction} \\ 
\noalign{\smallskip} \tableline \noalign{\smallskip} 
PG 1352+183  & 1.076 $\pm$ 0.106 & $-$0.07 $\pm$ 0.02 & & 2.69 $\pm$ 0.13 & 0.75 $\pm$ 0.04 & 1.41 $\pm$ 0.17 & & 0.0361 $\pm$ 0.0017 
                                & 0.92 $\pm$ 0.06 & 0.52 $\pm$ 0.04 & 0.37 $\pm$ 0.02 & 0.12 $\pm$ 0.01 \\
PG 1115+407  & 1.207 $\pm$ 0.184 & 0.05 $\pm$ 0.04    & & 1.85 $\pm$ 0.09 & 1.29 $\pm$ 0.06 & 0.82 $\pm$ 0.15 & & 0.0340 $\pm$ 0.0021  
                                & 0.87 $\pm$ 0.07 & 0.67 $\pm$ 0.06 & 0.44 $\pm$ 0.02 & 0.21 $\pm$ 0.02 \\
PG 1626+554  & 1.995 $\pm$ 0.279 & $-$0.20 $\pm$ 0.00 & & 3.43 $\pm$ 0.17 & 1.05 $\pm$ 0.05 & 1.50 $\pm$ 0.24  & & 0.0148 $\pm$ 0.0013 
                                & 1.17 $\pm$ 0.10 & 0.78 $\pm$ 0.08 & 0.52 $\pm$ 0.02 & 0.26 $\pm$ 0.03 \\
I Zw 1       & 10.55 $\pm$ 0.64  & $-$1.19 $\pm$ 0.00 & & 1.56 $\pm$ 0.00 & 10.1 $\pm$ 0.0 & 0.70 $\pm$ 0.05 &  & 0.0958 $\pm$ 0.0003
                                & 1.00 $\pm$ 0.00 & 1.00 $\pm$ 0.01 & 0.98 $\pm$ 0.01 & 1.01 $\pm$ 0.01 \\
PG 1114+445  & 1.904 $\pm$ 0.106 & $-$1.04 $\pm$ 0.00 & & 2.24 $\pm$ 0.06 & 1.45 $\pm$ 0.06 & 1.94 $\pm$ 0.11 & & 0.0138 $\pm$ 0.0008 
                                & 0.98 $\pm$ 0.08 & 0.79 $\pm$ 0.07 & 1.07 $\pm$ 0.02 & 0.42 $\pm$ 0.02 \\
3C 277.1     & 0.359 $\pm$ 0.015 & $-$0.24 $\pm$ 0.01 & & 1.78 $\pm$ 0.04 & 0.19 $\pm$ 0.01 &  1.24 $\pm$ 0.05 & & 0.0358 $\pm$ 0.0013  
                                & 0.66 $\pm$ 0.04 & 0.44 $\pm$ 0.03 & 0.23 $\pm$ 0.01 & 0.14 $\pm$ 0.01 \\
PG 1322+659  & 1.111 $\pm$ 0.142 & $-$0.20 $\pm$ 0.02 & & 1.83 $\pm$ 0.00 & 0.20 $\pm$ 0.01 & 1.96 $\pm$ 0.29 & & 0.0288 $\pm$ 0.0013 
                                & 0.91 $\pm$ 0.05 & 0.61 $\pm$ 0.04 & 1.05 $\pm$ 0.02 & 0.42 $\pm$ 0.02 \\
PG 0947+396  & 1.233 $\pm$ 0.061 & $-$0.06 $\pm$ 0.00 & & 1.19 $\pm$ 0.06 & 1.11 $\pm$ 0.06 & 1.27 $\pm$ 0.07 & & 0.0344 $\pm$ 0.0008 
                                & 0.82 $\pm$ 0.04 & 0.44 $\pm$ 0.03 & 0.31 $\pm$ 0.01 & 0.10 $\pm$ 0.01 \\
PG 1309+355  & 2.551 $\pm$ 0.183 & $-$0.63 $\pm$ 0.02 & & 3.73 $\pm$ 0.19 & 0.47 $\pm$ 0.02 & 1.12 $\pm$ 0.09 & & 0.0249 $\pm$ 0.0008 
                                & 0.84 $\pm$ 0.05 & 0.33 $\pm$ 0.05 & 0.82 $\pm$ 0.02 & 0.42 $\pm$ 0.02 \\
3C 323.1     & 1.330 $\pm$ 0.077 & $-$0.21 $\pm$ 0.02 & & 1.92 $\pm$ 0.00 & 0.02 $\pm$ 0.00 & 1.51 $\pm$ 0.09 & & 0.0215 $\pm$ 0.0010
                                 & 0.55 $\pm$ 0.08 & 0.30 $\pm$ 0.05 & 0.23 $\pm$ 0.01 & 0.11 $\pm$ 0.01 \\
B2 2201+31A & 4.345 $\pm$ 0.687 & $-$0.09 $\pm$ 0.02 & & 1.53 $\pm$ 0.04 & 12.4 $\pm$ 0.31 & 1.21 $\pm$ 0.23 & & 0.0164 $\pm$ 0.0008 
                                & 1.24 $\pm$ 0.06 & 0.76 $\pm$ 0.05 & 0.30 $\pm$ 0.08 & 0.39 $\pm$ 0.08 \\
QSO 0742+318 & 2.323 $\pm$ 0.232 & $-$0.73 $\pm$ 0.00 & & 20.9 $\pm$ 0.0 & 0.001 $\pm$ 0.000 & 1.43 $\pm$ 0.16 & & 0.0050 $\pm$ 0.0006 
                                & 0.70 $\pm$ 0.20 & 0.44 $\pm$ 0.11 & 0.11 $\pm$ 0.05 & 0.00            \\
3C 273       & 27.24 $\pm$ 0.12 & $-$0.14 $\pm$ 0.00 & & 2.23 $\pm$ 0.01 & 1.87 $\pm$ 0.09 & 0.75 $\pm$ 0.00 & & 0.0156 $\pm$ 0.0000  
                                & 1.14 $\pm$ 0.01 & 0.96 $\pm$ 0.01 & 0.75 $\pm$ 0.00 & 0.66 $\pm$ 0.00 \\
3C 334.0     & 0.731 $\pm$ 0.101 & 0.15 $\pm$ 0.01 & & 2.92 $\pm$ 0.15 & 0.01 $\pm$ 0.00 & 2.65 $\pm$ 0.42 & & 0.0532 $\pm$ 0.0019 
                                & 0.87 $\pm$ 0.05 & 0.25 $\pm$ 0.01 & 0.31 $\pm$ 0.01 & 0.27 $\pm$ 0.01 \\
\noalign{\smallskip} \tableline \noalign{\smallskip} 
\multicolumn{13}{c}{With correction for the SMC-like intrinsic
 extinction of $E_{B-V} = 0.09$} \\ 
\noalign{\smallskip} \tableline \noalign{\smallskip} 
I Zw 1        & 13.43 $\pm$ 0.82  & $-$0.47 $\pm$ 0.00 & & 0.72 $\pm$ 0.02 & 440 & 0.10 $\pm$ 0.01 & & 0.1221 $\pm$ 0.0003
                               & 0.99 $\pm$ 0.03 & 1.01 $\pm$ 0.01 & 0.65 $\pm$ 0.01 & 0.71 $\pm$ 0.01 \\   
PG 1114+445  & 2.417 $\pm$ 0.134 & $-$0.27 $\pm$ 0.01 & & 1.57 $\pm$ 0.04 & 1.70 $\pm$ 0.10 & 1.84 $\pm$ 0.85 & & 0.0221 $\pm$ 0.0012  
                                & 0.82 $\pm$ 0.07 & 0.60 $\pm$ 0.05 & 0.51 $\pm$ 0.01 & 0.15 $\pm$ 0.01 \\
\enddata
\tablenotetext{a}{Flux density at 5700 $\mbox{\AA}$ in units of $10^{-26}$ ergs s$^{-1}$ cm$^{-2}$ Hz$^{-1}$, corresponding to a flux of $ 5.260 \times 10^{-12}$ ergs s$^{-1}$ cm$^{-2}$.}
\tablenotetext{b}{Relative strength of the Balmer continuum at the Balmer edge, which is defined as $h_{BE} = aB_{\nu}(T_e)(1-e^{-\tau_{BE}})$.}
\end{deluxetable}

\begin{deluxetable}{lcccccc}
\tablecaption{FWHMs of \ion{Mg}{2}, H$\alpha$, and [\ion{O}{3}] \label{tabFWHM}}
\tablewidth{0pt}
\tablehead{
\colhead{Quasars} & \colhead{\ion{Mg}{2} FWHM} & \colhead{H$\alpha$ FWHM} & \colhead{[\ion{O}{3}] FWHM} & &
\colhead{\ion{Mg}{2} ratio\tablenotemark{a}} & \colhead{H$\alpha$ ratio\tablenotemark{b}}\\
\colhead{}        & \colhead{(km s$^{-1}$)}    & \colhead{(km s$^{-1}$)}  & \colhead{(km s$^{-1}$)}     & & 
\colhead{} & \colhead{} 
}
\startdata
PG 1352+183  & 3440 $\pm$ 290 & 2660 $\pm$ 70  & 1970 $\pm$ 210 & &  1.05 $\pm$ 0.14 & 1.04 $\pm$ 0.02 \\
PG 1115+407  & 2460 $\pm$ 240 & -              &  680 $\pm$ 110 & &  1.02 $\pm$ 0.23 & - \\
PG 1626+554  & 4160 $\pm$ 310 & 4100 $\pm$ 40  & 1510 $\pm$ 180 & &  1.07 $\pm$ 0.14 & 1.06 $\pm$ 0.02 \\
IZw1         & 1660 $\pm$ 10  & 1490 $\pm$ 20  &  640 $\pm$ 60  & &  1.06 $\pm$ 0.01 & 1.14 $\pm$ 0.02 \\
PG 1114+445  & 4620 $\pm$ 370 & 5030 $\pm$ 30  & 1190 $\pm$ 50  & &  1.06 $\pm$ 0.12 & 1.07 $\pm$ 0.01 \\
3C 277.1     & 3380 $\pm$ 130 & 3110 $\pm$ 30  &  610 $\pm$ 10  & &  1.22 $\pm$ 0.09 & 1.07 $\pm$ 0.01 \\
PG 1322+659  & 2700 $\pm$ 170 & 3260 $\pm$ 30  &  690 $\pm$ 50  & &  1.15 $\pm$ 0.13 & 1.10 $\pm$ 0.02 \\
PG 0947+396  & 4090 $\pm$ 360 & 4750 $\pm$ 20  &  830 $\pm$ 40  & &  1.02 $\pm$ 0.23 & 1.06 $\pm$ 0.01 \\
PG 1309+355  & 3650 $\pm$ 250 & 3800 $\pm$ 50  & 1300 $\pm$ 60  & &  1.08 $\pm$ 0.12 & 1.07 $\pm$ 0.02 \\
3C 323.1     & 6270 $\pm$ 320 & 6690 $\pm$ 50  &  760 $\pm$ 10  & & 1.11 $\pm$ 0.12 & 1.06 $\pm$ 0.02 \\
B2 2201+31A  & 3710 $\pm$ 290 & 4380 $\pm$ 150 & 1650 $\pm$ 550 & & 1.00 $\pm$ 0.09 & 1.05 $\pm$ 0.11 \\
QSO 0742+318 & 7680 $\pm$ 770 & -              & 1170 $\pm$ 70  & & 1.02 $\pm$ 0.26 & - \\
3C 273       & 3400 $\pm$ 20  & 4480 $\pm$ 10  & 2070 $\pm$ 30  & & 1.06 $\pm$ 0.01 & 1.09 $\pm$ 0.00 \\
3C 334.0     & 4950 $\pm$ 420 & 7390 $\pm$ 20  & 910 $\pm$ 10   & & 1.11 $\pm$ 0.20 & 1.04 $\pm$ 0.01 \\
\enddata
\tablecomments{FWHMs were measured by applying the single Gaussian component 
to the spectrum where the power-law and Balmer continua and \ion{Fe}{2} 
emission lines have been subtracted.}
\tablenotetext{a}{The ratio between \ion{Mg}{2} flux measured by fitting the two Gaussian components and 
that by fitting a single Gaussian component.}
\tablenotetext{b}{The ratio between H$\alpha$ flux measured by fitting the two Gaussian components and 
that by fitting a single Gaussian component.}
\end{deluxetable}

\begin{deluxetable}{lcccccccc}
\tabletypesize{\scriptsize}
\rotate
\tablecaption{Observed Emission-Line and Continuum Fluxes\label{lines1}}
\tablewidth{0pt}
\tablehead{
\colhead{Line or continuum} & \colhead{PG 1352+183} & \colhead{PG
 1115+407} & \colhead{PG 1626+554} & \colhead{I Zw 1} &
 \colhead{I Zw 1\tablenotemark{a}} & \colhead{PG 1114+445} & \colhead{PG 1114+445\tablenotemark{a}} & \colhead{3C 277.1} 
}
\startdata
Ly$\alpha$ $\lambda$1216 & 17.0 $\pm$ 1.3 & 14.4 $\pm$ 1.3 & 12.4 $\pm$ 0.8 & 9.18 $\pm$ 0.46 
 & 31.2 $\pm$ 1.6 & 3.12 $\pm$ 0.16 & 12.6 $\pm$ 0.7 & - \\  
N V $\lambda$1240            & 1.92 $\pm$ 0.41 & 2.78 $\pm$ 0.38 & 3.60 $\pm$ 0.50 & 2.15 $\pm$ 0.11 
 & 6.74 $\pm$ 0.35 & 0.24 $\pm$ 0.06 & 0.30 $\pm$ 0.06 & - \\  
\ion{O}{1} $\lambda$1304     & - & -  & - & 0.40 $\pm$ 0.03 & 1.40 $\pm$ 0.09 
 & - & - & - \\  
\ion{Si}{4} $\lambda$1400    & 1.68 $\pm$ 0.46 & - & 1.89 $\pm$ 0.37 & 1.03 $\pm$ 0.07 
 & 2.44 $\pm$ 0.16 & 0.48 $\pm$ 0.07 & 1.49 $\pm$ 0.20 & 1.73 $\pm$ 0.24 \\  
\ion{C}{4} $\lambda$1549     & 8.92 $\pm$ 0.73 & 8.35 $\pm$ 0.90 & 6.33 $\pm$ 0.42 & 1.77 $\pm$ 0.28 
 & 3.82 $\pm$ 0.53 & 1.68 $\pm$ 0.10 & 3.74 $\pm$ 0.21 & 9.52 $\pm$ 0.50 \\  
\ion{Si}{3}] $\lambda$1892   & 2.94 $\pm$ 0.30 & 6.09 $\pm$ 0.44 & 1.45 $\pm$ 0.12 & 0.88 $\pm$ 0.05 
 & 1.52 $\pm$ 0.08 & 0.61 $\pm$ 0.06 & 0.99 $\pm$ 0.11 & 1.04 $\pm$ 0.11 \\ 
\ion{C}{3}] $\lambda$1909    & 1.62 $\pm$ 0.15 & 1.90 $\pm$ 0.16 & 0.82 $\pm$ 0.07 & 0.79 $\pm$ 0.05 
 & 1.38 $\pm$ 0.10 & 0.37 $\pm$ 0.06 & 0.72 $\pm$ 0.10 & 0.84 $\pm$ 0.07 \\  
\ion{Mg}{2} $\lambda$2798    & 1.45 $\pm$ 0.12 & 1.34 $\pm$ 0.20 & 1.23 $\pm$ 0.09 & 1.27 $\pm$ 0.06 
 & 1.66 $\pm$ 0.08 & 0.65 $\pm$ 0.05 & 0.86 $\pm$ 0.07 & 1.11 $\pm$ 0.07 \\  
\ion{Fe}{2}($U1$) & 3.96 $\pm$ 0.33 & 4.23 $\pm$ 0.46 & 1.71 $\pm$ 0.17 & 4.89 $\pm$ 0.24 & 8.06 $\pm$ 0.39 
 & 1.28 $\pm$ 0.10 & 2.06 $\pm$ 0.17 & 2.91 $\pm$ 0.19 \\  
\ion{Fe}{2}($U2$) & 2.89 $\pm$ 0.27 & 2.95 $\pm$ 0.47 & 1.60 $\pm$ 0.16 & 3.91 $\pm$ 0.20 & 5.45 $\pm$ 0.27 
 & 1.00 $\pm$ 0.10 & 1.34 $\pm$ 0.13 & 1.53 $\pm$ 0.13 \\  
\ion{Fe}{2}($U3$) & 1.51 $\pm$ 0.16 & 2.07 $\pm$ 0.35 & 0.98 $\pm$ 0.11 & 3.59 $\pm$ 0.19 & 4.50 $\pm$ 0.23 
 & 0.74 $\pm$ 0.07 & 0.89 $\pm$ 0.09 & 0.94 $\pm$ 0.07 \\  
\ion{Fe}{2}($O1$) & 0.96 $\pm$ 0.08 & 1.23 $\pm$ 0.13 & 0.59 $\pm$ 0.05 & 3.12 $\pm$ 0.20 & 2.65 $\pm$ 0.18 
 & 0.90 $\pm$ 0.06 & 0.69 $\pm$ 0.05 & 0.43 $\pm$ 0.03 \\  
\ion{Fe}{2}($O2$) & 0.29 $\pm$ 0.05 & 0.55 $\pm$ 0.10 & 0.27 $\pm$ 0.04 & 3.01 $\pm$ 0.17 & 2.68 $\pm$ 0.16 
 & 0.33 $\pm$ 0.02 & 0.19 $\pm$ 0.02 & 0.25 $\pm$ 0.02 \\  
H$\delta$ $\lambda$4102            & - & - & 0.25 $\pm$ 0.04 & 0.14 $\pm$ 0.01 
 & 0.15 $\pm$ 0.02 & - & - & 0.23 $\pm$ 0.02 \\  
H$\gamma$ $\lambda$4340            & - & 0.35 $\pm$ 0.04 & 0.39 $\pm$ 0.03 & 0.44 $\pm$ 0.03 
 & 0.43 $\pm$ 0.03 & 0.06 $\pm$ 0.01 & - & 0.42 $\pm$ 0.03 \\   
H$\beta$ $\lambda$4861             & 1.00 $\pm$ 0.05 & 1.00 $\pm$ 0.05 & 1.00 $\pm$ 0.04 & 1.00 $\pm$ 0.04 
 & 1.00 $\pm$ 0.04 & 1.00 $\pm$ 0.04 & 1.00 $\pm$ 0.04 & 1.00 $\pm$ 0.04 \\  
$[$\ion{O}{3}] $\lambda$5007       & 0.31 $\pm$ 0.02 & 0.13 $\pm$ 0.01 & 0.13 $\pm$ 0.01 & 0.13 $\pm$ 0.01 
 & 0.14 $\pm$ 0.01 & 0.28 $\pm$ 0.02 & 0.29 $\pm$ 0.02 & 0.59 $\pm$ 0.03 \\  
H$\alpha$ $\lambda$6563      & 2.48 $\pm$ 0.15 & - & 3.27 $\pm$ 0.20 & 4.48 $\pm$ 0.26 
 & 4.23 $\pm$ 0.24 & 2.50 $\pm$ 0.14 & 2.38 $\pm$ 0.14 & 3.02 $\pm$ 0.17 \\  
$\lambda$$F_{\lambda}$(1450$\mbox{\AA}$) & 215 $\pm$ 34 & 309 $\pm$ 46 & 155 $\pm$ 27 
 & 97.5 $\pm$ 6.1 & 238 $\pm$ 15 & 42.3 $\pm$ 6.5 & 104 $\pm$ 16 & 139 $\pm$ 19 \\
Bac & 30.6 $\pm$ 5.8 & 21.5 $\pm$ 4.7 & 29.9 $\pm$ 5.1 & 23.4 $\pm$ 1.0 & 2.25 $\pm$ 0.09 
 & 25.6 $\pm$ 3.5 & 19.6 $\pm$ 7.7 & 16.1 $\pm$ 2.2 \\
\noalign{\smallskip} \tableline \noalign{\smallskip}
H$\beta$ flux$^{b}$ & 0.76 $\pm$ 0.04 & 0.61 $\pm$ 0.03 & 1.65 $\pm$
 0.07 & 4.79 $\pm$ 0.19 & 6.08 $\pm$ 0.24 & 2.11 $\pm$ 0.08 & 2.67 $\pm$
 0.11 & 0.24 $\pm$ 0.01 \\  
\noalign{\smallskip} \tableline \noalign{\smallskip}
$\lambda L_{3000}^{\ \ \ \ \ c}$ & 6.38 $\pm$ 0.66 & 7.04 $\pm$ 0.74 & 8.20 $\pm$ 0.84 & 5.43 $\pm$ 0.10 & 
 8.72 $\pm$ 0.16 & 7.81 $\pm$ 0.78 & 12.6 $\pm$ 1.3 & 6.39 $\pm$ 0.58 \\  
\noalign{\smallskip} \tableline \noalign{\smallskip}
\tablehead{
\colhead{Line or continuum} & \colhead{PG 1322+659} & \colhead{PG 0947+396} & \colhead{PG 1309+355} & \colhead{3C 323.1} &
 \colhead{B2 2201+31A} & \colhead{QSO 0742+318} & \colhead{3C 273} & \colhead{3C 334.0} 
}
\tablebreak
Ly$\alpha$ $\lambda$1216     & 12.8 $\pm$ 0.7 & 12.0 $\pm$ 0.7 & 7.47 $\pm$ 0.44 & 17.9 $\pm$ 1.1 
 & 7.99 $\pm$ 1.10 & 9.94 $\pm$ 0.78 & - & 6.67 $\pm$ 0.40 \\
 \ion{N}{5} $\lambda$1240    & 1.16 $\pm$ 0.09 & 1.70 $\pm$ 0.19 & - & 3.05 $\pm$ 0.66 
 & 1.64 $\pm$ 0.33 & 2.68 $\pm$ 0.26 & - & 2.92 $\pm$ 0.66 \\
\ion{O}{1} $\lambda$1304     & - & - & - & - & - & - & - & - \\
\ion{Si}{4} $\lambda$1400    & 0.57 $\pm$ 0.10 & 0.86 $\pm$ 0.11 & - & 1.48 $\pm$ 0.39 
 & 1.15 $\pm$ 0.20 & 0.43 $\pm$ 0.08 & - & 0.61 $\pm$ 0.08 \\
\ion{C}{4} $\lambda$1549     & 6.25 $\pm$ 0.36 & 7.38 $\pm$ 0.45 & 3.39 $\pm$ 0.23 & 13.8 $\pm$ 0.8 
 & 5.95 $\pm$ 1.11 & 6.43 $\pm$ 0.49 & 2.83 $\pm$ 0.13 & 6.29 $\pm$ 0.50 \\
\ion{Si}{3}] $\lambda$1892   & 2.07 $\pm$ 0.13 & 2.67 $\pm$ 0.19 & 0.93 $\pm$ 0.23 & 0.94 $\pm$ 0.24 
 & 3.90 $\pm$ 0.65 & - & - & 2.81 $\pm$ 0.30 \\
\ion{C}{3}] $\lambda$1909    & 1.15 $\pm$ 0.06 & 1.07 $\pm$ 0.08 & 1.08 $\pm$ 0.25 & 1.05 $\pm$ 0.11 
 & 1.78 $\pm$ 0.28 & 1.85 $\pm$ 0.19 & 2.12 $\pm$ 0.10 & 0.67 $\pm$ 0.16 \\ 
\ion{Mg}{2} $\lambda$2798    & 0.76 $\pm$ 0.08 & 1.05 $\pm$ 0.08 & 1.16 $\pm$ 0.08 & 2.05 $\pm$ 0.18 
 & 1.08 $\pm$ 0.15 & 1.36 $\pm$ 0.16 & 0.91 $\pm$ 0.05 & 0.94 $\pm$ 0.08 \\
\ion{Fe}{2}($U1$)              & 2.21 $\pm$ 0.22 & 4.00 $\pm$ 0.26 & 3.01 $\pm$ 0.19 & 3.18 $\pm$ 0.29 
 & 2.61 $\pm$ 0.43 & 1.62 $\pm$ 0.22 & 1.95 $\pm$ 0.13 & 3.52 $\pm$ 0.22 \\
\ion{Fe}{2}($U2$)              & 1.60 $\pm$ 0.17 & 2.62 $\pm$ 0.19 & 2.02 $\pm$ 0.16 & 1.39 $\pm$ 0.23 
 & 2.58 $\pm$ 0.37 & 0.89 $\pm$ 0.26 & 1.77 $\pm$ 0.10 & 2.43 $\pm$ 0.18 \\
\ion{Fe}{2}($U3$)              & 0.98 $\pm$ 0.11 & 1.27 $\pm$ 0.13 & 0.73 $\pm$ 0.11 & 0.69 $\pm$ 0.12 
 & 1.44 $\pm$ 0.23 & 0.52 $\pm$ 0.14 & 1.36 $\pm$ 0.07 & 0.63 $\pm$ 0.04 \\
\ion{Fe}{2}($O1$)              & 1.53 $\pm$ 0.14 & 0.82 $\pm$ 0.06 & 1.60 $\pm$ 0.11 & 0.48 $\pm$ 0.04 
 & 0.50 $\pm$ 0.14 & 0.11 $\pm$ 0.05 & 0.96 $\pm$ 0.06 & 0.71 $\pm$ 0.05 \\
\ion{Fe}{2}($O2$)              & 0.56 $\pm$ 0.05 & 0.23 $\pm$ 0.05 & 0.76 $\pm$ 0.05 & 0.21 $\pm$ 0.02 
 & 0.60 $\pm$ 0.14 & - & 0.77 $\pm$ 0.04 & 0.57 $\pm$ 0.04 \\
H$\delta$ $\lambda$4102      & 0.19 $\pm$ 0.02 & - & - & 0.27 $\pm$ 0.03 
 & - & - & 0.19 $\pm$ 0.01 & - \\
H$\gamma$ $\lambda$4340      & 0.24 $\pm$ 0.02 & 0.43 $\pm$ 0.03 & 0.14 $\pm$ 0.03 & 0.35 $\pm$ 0.03
 & - & 0.30 $\pm$ 0.06 & 0.34 $\pm$ 0.02 & 0.34 $\pm$ 0.02 \\
H$\beta$ $\lambda$4861       & 1.00 $\pm$ 0.04 & 1.00 $\pm$ 0.04 & 1.00 $\pm$ 0.04 & 1.00 $\pm$ 0.04 
 & 1.00 $\pm$ 0.12 & 1.00 $\pm$ 0.07 & 1.00 $\pm$ 0.04 & 1.00 $\pm$ 0.04 \\
$[$\ion{O}{3}] $\lambda$5007 & 0.12 $\pm$ 0.01 & 0.17 $\pm$ 0.01 & 0.49 $\pm$ 0.03 & 0.40 $\pm$ 0.02 
 & - & 0.80 $\pm$ 0.08 & 0.14 $\pm$ 0.01 & 0.63 $\pm$ 0.04 \\ 
H$\alpha$ $\lambda$6563  & 1.91 $\pm$ 0.11 & 2.99 $\pm$ 0.17 & 2.53 $\pm$ 0.15 & 3.84 $\pm$ 0.22 
 & 3.89 $\pm$ 0.53 & - & 3.42 $\pm$ 0.19 & 2.06 $\pm$ 0.12 \\ 
$\lambda$$F_{\lambda}$(1450$\mbox{\AA}$)  & 136 $\pm$ 12 & 170 $\pm$ 16 & 132 $\pm$ 16 
 & 141 $\pm$ 25 & 229 $\pm$ 35 & 81.9 $\pm$ 8.4 & 197 $\pm$ 10 & 108 $\pm$ 12 \\
Bac     & 24.8 $\pm$ 1.6 & 14.4 $\pm$ 4.2 & 32.7 $\pm$ 5.4 & 18.9 $\pm$ 0.8 & 23.2 $\pm$ 4.0 
 & 20.8 $\pm$ 1.5 & 17.6 $\pm$ 1.1 & 20.6 $\pm$ 2.8 \\
\noalign{\smallskip} \tableline \noalign{\smallskip}
H$\beta$ flux$^b$ & 0.99 $\pm$ 0.04 & 0.94 $\pm$ 0.04 & 1.36 $\pm$ 0.05 
 & 0.95 $\pm$ 0.04 & 2.30 $\pm$ 0.28 & 1.12 $\pm$ 0.08 & 16.5 $\pm$ 0.7 
 & 0.80 $\pm$ 0.03 \\
\noalign{\smallskip} \tableline \noalign{\smallskip}
$\lambda L_{3000}^{\ \ \ \ \ c}$ & 8.01 $\pm$ 0.58 & 12.2 $\pm$ 1.1 & 14.5 $\pm$ 1.4 & 18.3 $\pm$ 2.0 & 
 74.0 $\pm$ 5.7 & 49.3 $\pm$ 4.9 & 142 $\pm$ 1 & 51.3 $\pm$ 9.0 \\ 
\enddata
\tablecomments{Fluxes are given in the observer frame relative to 
the H$\beta$ flux. \ion{Fe}{2}($U1$), \ion{Fe}{2}($U2$), .., and 
\ion{Fe}{2}($O2$) are the fluxes in the wavelength bands of 
$U1$ (2200$-$2660 \AA), $U2$ (2660$-$3000 \AA), $U3$ (3000$-$3500 \AA), 
$O1$ (4400$-$4700 \AA), and $O2$ (5100$-$5600 \AA), respectively. 
$\lambda F_\lambda(1450\AA)$ is the flux at 1450 \AA, and Bac represents 
the total flux of the Balmer continuum.}
\tablenotetext{a}{The SMC-like intrinsic extinction of E$_{B-V}$ = 0.09 
has been taken into account.}
\tablenotetext{b}{The flux of H$\beta$ is given in units of 
$10^{-13}$ ergs s$^{-1}$ cm$^{-2}$. }
\tablenotetext{c}{$\lambda L_{3000}$, the 
luminosity at 3000 \AA\ is given in units of 10$^{37}$ W.}
\end{deluxetable}

\begin{deluxetable}{lccccccc}
\tabletypesize{\scriptsize}
\rotate
\tablecaption{Comparison with the Work by \citet{wnw} \label{compWNW}}
\tablewidth{0pt}
\tablehead{
\colhead{Line or continuum} & \multicolumn{3}{c}{QSO 0742+318} &
 \colhead{} & \multicolumn{3}{c}{3C 273} \\
\cline{2-4} \cline{6-8} \\
\colhead{$X$} &  \colhead{$F_{X}/F_{H\beta}$(WNW)\tablenotemark{a}} &
 \colhead{$F_{X}/F_{H\beta}$(Ours)\tablenotemark{b}} &
 \colhead{$F_{X}$(WNW)/$F_{X}$(Ours)\tablenotemark{c}} & \colhead{} & \colhead{$F_{X}/F_{H\beta}$(WNW)\tablenotemark{a}} & \colhead{$F_{X}/F_{H\beta}$(Ours)\tablenotemark{b}} & \colhead{$F_{X}$(WNW)/$F_{X}$(Ours)\tablenotemark{c}}
}
\startdata
\ion{Mg}{2} $\lambda2798$   & 1.45 $\pm$ 0.29 &  1.36 $\pm$ 0.16 & 1.07 $\pm$ 0.24 
                            & & 1.35 $\pm$ 0.27 &  0.91 $\pm$ 0.05 & 1.48 $\pm$ 0.31 \\
\ion{Fe}{2}(2000$-$3000 \AA) & 5.05            &  2.67 $\pm$ 0.36 & 1.89          
                            &   & 9.8             &  3.96 $\pm$ 0.17 & 2.47            \\
\ion{Fe}{2}(3000$-$3500 \AA)    & 0.51            &  0.52 $\pm$ 0.14 & 0.98  
                            &  & 1.1             &  1.36 $\pm$ 0.07 & 0.81            \\
\ion{Fe}{2}(3500$-$6000 \AA)    & 0.65            &                  &  
                            &  & 2.89            &  3.45 $\pm$ 0.14 & 0.84  \\
Bac                         & 7.78            &  20.8 $\pm$ 1.5 & 0.37
                            &  & 7.6             &  17.6 $\pm$ 1.1 & 0.43 \\
H$\beta$                    & 1.00            &  1.00            & 1.17 $\pm$ 0.14 
                            &  & 1.00            &  1.00 & 1.06 $\pm$ 0.07 \\
\enddata
\tablenotetext{a}{Ratio of $X$-flux relative to $H\beta$ 
by \citet{wnw}.}
\tablenotetext{b}{Ratio of $X$-flux relative to $H\beta$ 
by this work.}
\tablenotetext{c}{Ratio of $X$-flux by \citet{wnw} relative 
to that by this work.}
\end{deluxetable}

\begin{deluxetable}{cllcrclrclrc}
\tabletypesize{\scriptsize}
\tablecaption{Results of Linear Regression ($Y = A + BX$) and Correlation 
Coefficient \label{rank}}
\tablewidth{0pt}
\tablehead{
\colhead{Panel} & \colhead{$X$\tablenotemark{a}} &
 \colhead{$Y$\tablenotemark{a}} & $N$ & \multicolumn{3}{c}{$A$} & \multicolumn{3}{c}{$B$} &
 \colhead{$r$} & \colhead{C.L.(\%)\tablenotemark{b}} 
}
\startdata
($a$)   & \ion{Mg}{2} FWHM & H$\alpha$ FWHM & 12 & $-$572   & $\pm$ & 825   &    1.26   & $\pm$ & 0.21   & 0.886    & 99.99 \\
($b$)   & \ion{Mg}{2} FWHM & $M_{B}$        & 14 & $-$22.3  & $\pm$ & 0.9   & $-$0.439  & $\pm$ & 0.213  & $-$0.510 & 93.78 \\
($c$)   & Bac/H$\beta$     & \ion{Fe}{2}($O1$)/\ion{Fe}{2}($U1$) 
                                            & 14 & $-$0.0185 & $\pm$ & 0.247   & 0.0161 & $\pm$ & 0.0105 & 0.404 & 84.77 \\
($d$)   & Bac/H$\beta$ & \ion{Fe}{2}($O2$)/H$\beta$ 
                                            & 13 & 0.436 & $\pm$ & 0.926 &    0.0091 & $\pm$ & 0.0392 & 0.07 & 18.02 \\
($e$)   &  $\Gamma$    & $M_{B}$            & 14 & $-$29.7  & $\pm$ & 1.7   & 2.36      & $\pm$ & 0.68   & 0.709    & 99.55 \\
($f$)   & $\Gamma$     & \ion{Fe}{2}($U1$)/\ion{Mg}{2} 
                                            & 14 & 0.507    & $\pm$ & 1.423 &  0.860    & $\pm$ & 0.584  & 0.391    & 83.33 \\
($g$)   & $\Gamma$    & \ion{Fe}{2}($O1$)/\ion{Mg}{2} 
                                            & 14 & $-$1.56  & $\pm$ & 0.95  &  1.04     & $\pm$ & 0.39   & 0.609    & 97.91 \\
($h$)   & $\Gamma$    & \ion{Fe}{2}($O1$)/\ion{Fe}{2}($U1$) 
                                            & 14 & $-$0.276 & $\pm$ & 0.340 &  0.260    & $\pm$ & 0.139  & 0.473    & 91.25 \\
($i$)   & \ion{Mg}{2} FWHM & $\Gamma$       & 14 & 3.23     & $\pm$ & 0.19  & $-$0.205  & $\pm$ & 0.046  & $-$0.792 & 99.93 \\
($j$)   & \ion{Mg}{2} FWHM & \ion{Fe}{2}($U1$)/\ion{Mg}{2} 
                                            & 14 & 3.99     & $\pm$ & 0.55  & $-$0.353  & $\pm$ & 0.128  & $-$0.622 & 98.25 \\
($k$)   & \ion{Mg}{2} FWHM & \ion{Fe}{2}($O1$)/\ion{Mg}{2} 
                                            & 14 & 2.13     & $\pm$ & 0.40  & $-$0.300  & $\pm$ & 0.093  & $-$0.682 & 99.27 \\
($l$)   & \ion{Mg}{2} FWHM & \ion{Fe}{2}($O1$)/\ion{Fe}{2}($U1$) 
                                            & 14 & 0.659    & $\pm$ & 0.147 & $-$0.0769 & $\pm$ & 0.00345 & $-$0.542 & 95.46 \\
($m$)   & $\log M_{BH}$ & $\Gamma$          & 14 & 3.34     & $\pm$ & 0.15  & $-$0.751  & $\pm$ & 0.111   & $-$0.889 & 99.99 \\
($n$)   & $\log M_{BH}$ & \ion{Fe}{2}($U1$)/\ion{Mg}{2} 
                                            & 14 & 3.84     & $\pm$ & 0.59  & $-$1.02   & $\pm$ & 0.45    & $-$0.549 & 95.80 \\
($o$)   & $\log M_{BH}$ & \ion{Fe}{2}($O1$)/\ion{Mg}{2} 
                                            & 14 & 2.17     & $\pm$ & 0.40  & $-$0.995   & $\pm$ & 0.300  & $-$0.691 & 99.38 \\
($p$)   & $\log M_{BH}$ & \ion{Fe}{2}($O1$)/\ion{Fe}{2}($U1$) 
                                            & 14 & 0.650    & $\pm$ & 0.151 & $-$0.242  & $\pm$ & 0.114   & $-$0.521 & 94.39 \\
($q$)   & [\ion{O}{3}]/H$\beta$ & \ion{Mg}{2} FWHM 
                                            & 13 & 2.44     & $\pm$ & 0.61  & 4.79      & $\pm$ & 1.52    & 0.690    & 99.09 \\
($r$)   & [\ion{O}{3}]/H$\beta$ & \ion{Fe}{2}($U1$)/\ion{Mg}{2} 
                                            & 13 & 2.93     & $\pm$ & 0.46  & $-$1.02   & $\pm$ & 1.15    & $-$0.258 & 60.55 \\
($s$)   &  [\ion{O}{3}]/H$\beta$ & \ion{Fe}{2}($O1$)/\ion{Mg}{2} 
                                            & 13 & 1.50     & $\pm$ & 0.30  & $-$1.61   & $\pm$ & 0.76    & $-$0.538 & 94.19 \\
($t$)   & [\ion{O}{3}]/H$\beta$ & \ion{Fe}{2}($O1$)/\ion{Fe}{2}($U1$) 
                                            & 13 &  0.545   & $\pm$ & 0.096 & $-$0.550  & $\pm$ & 0.239   & $-$0.570 & 95.80 \\ 
($u$)   & [\ion{O}{3}] FWHM & \ion{Fe}{2}($O1$)/[\ion{O}{3}] 
                                            & 13 & 9.74     & $\pm$ & 4.63   & $-$3.58   & $\pm$ & 3.85    & $-$0.270 & 62.79 \\
($v$)   & \ion{Mg}{2} FWHM & [\ion{O}{3}] FWHM 
                                            & 13 & 862      & $\pm$ & 458   & 39      & $\pm$ & 107    & 0.103    & 27.45 \\
($w$)   & \ion{Fe}{2}($U1$)/\ion{Mg}{2} & \ion{Fe}{2}($O1$)/\ion{Fe}{2}($U1$)  
                                            & 14 & 0.241    & $\pm$ & 0.193 & 0.0424    & $\pm$ & 0.0711  & 0.170    & 43.79 \\
($x$)   & \ion{Fe}{2}($O1$)/\ion{Mg}{2} & \ion{Fe}{2}($O1$)/\ion{Fe}{2}($U1$) 
                                            & 14 & 0.0830   & $\pm$ & 0.0487 & 0.287    & $\pm$ & 0.043   & 0.888    & 99.99 \\
\enddata
\tablenotetext{a}{FWHM in units of 1000 km s$^{-1}$.}
\tablenotetext{b}{Confidence level in percentage.}
\end{deluxetable}

\clearpage

\clearpage

\clearpage

\clearpage


\clearpage

\clearpage
\plotone{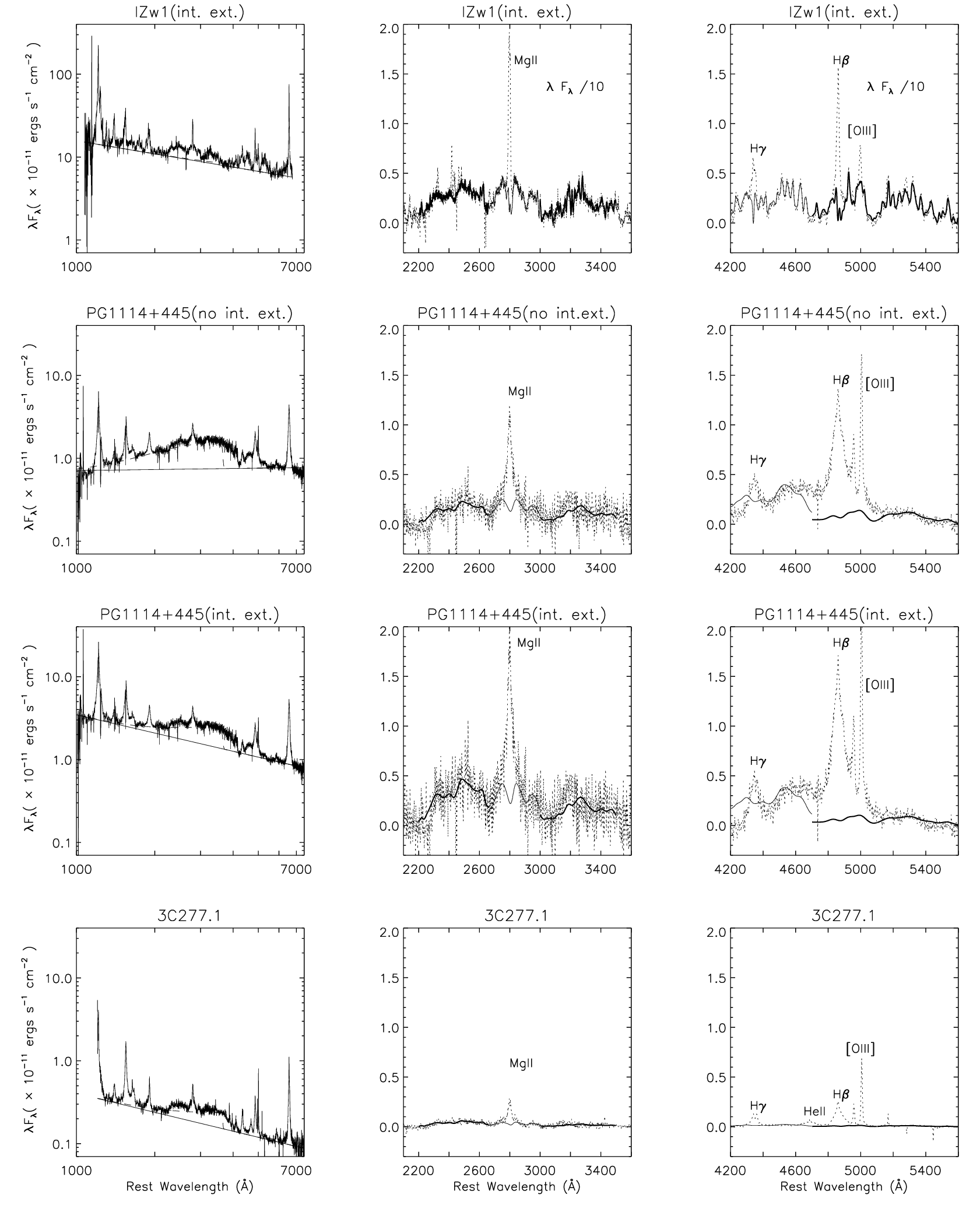}
\centerline{\label{qso2}Fig. 8. --- Continued.} 
\clearpage
\plotone{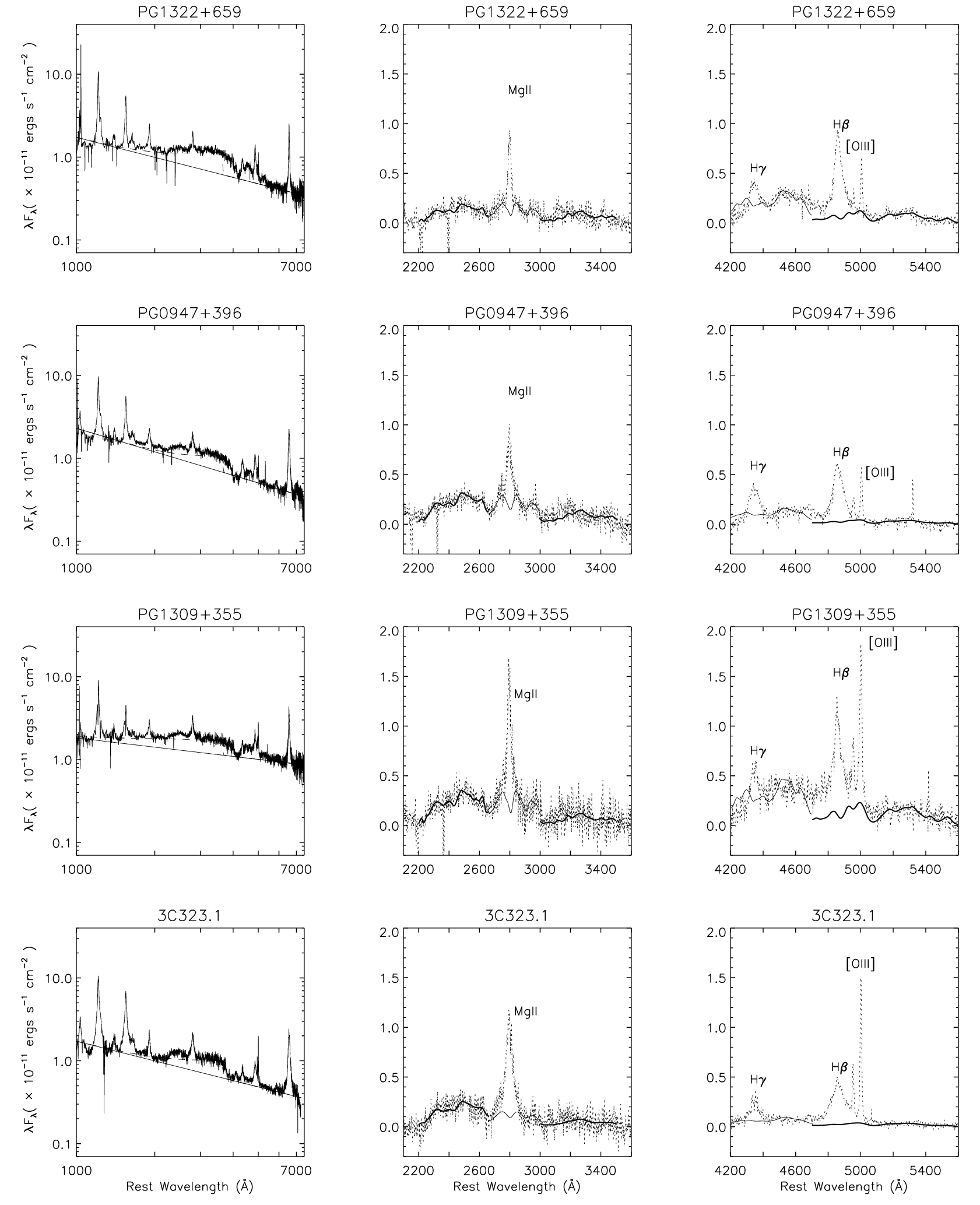}
\centerline{\label{qso3}Fig. 8. --- Continued.} 
\clearpage
\plotone{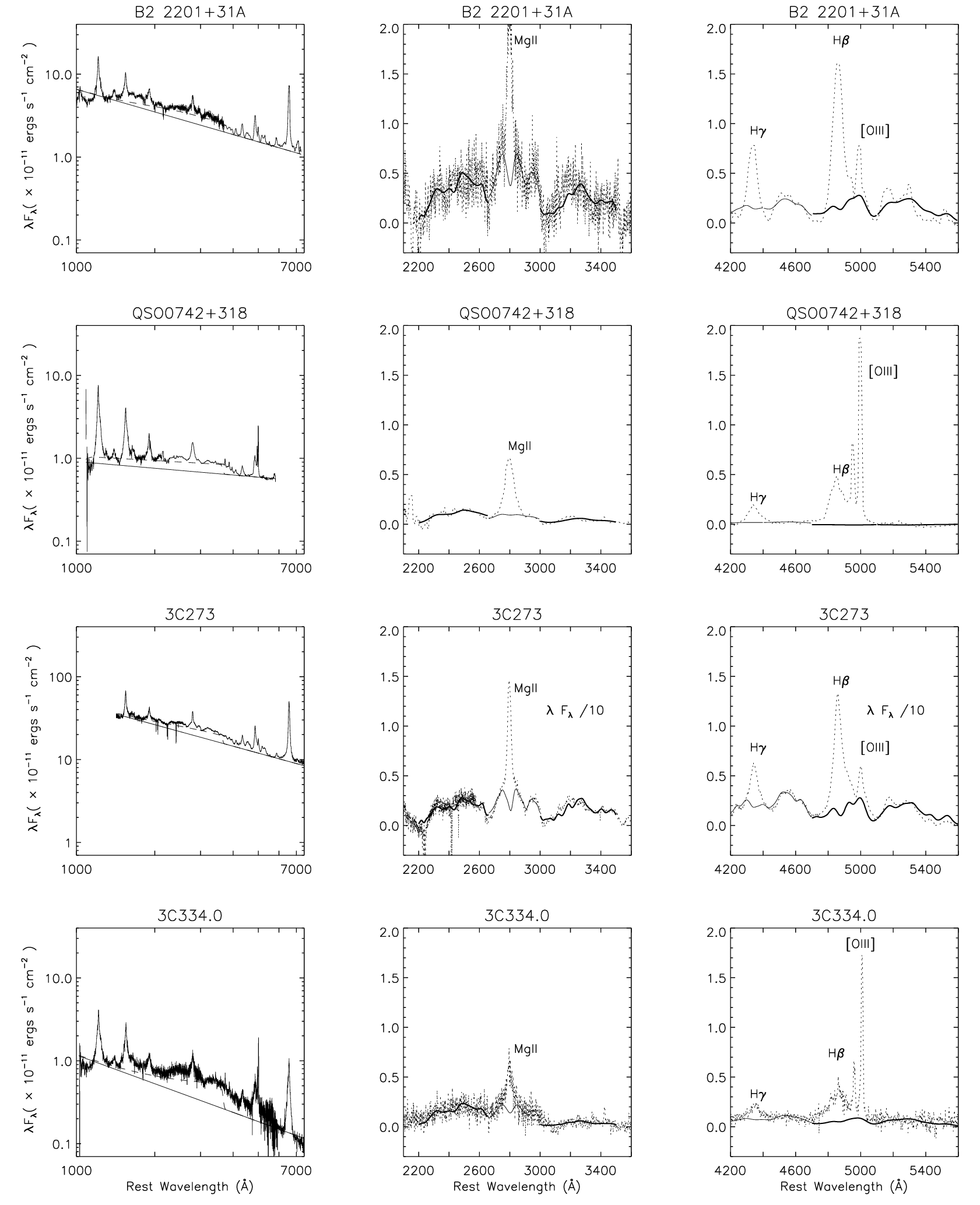} 
\centerline{\label{qso4}Fig. 8. --- Continued.} 
\clearpage

\end{document}